\documentclass[12pt,english,american,preprint]{elsarticle}
\usepackage[LGR,T1]{fontenc}
\usepackage[latin9]{inputenc}
\usepackage{color}
\usepackage{textcomp}
\usepackage{mathrsfs}
\usepackage{amsmath}
\usepackage{amssymb}
\usepackage{graphicx}

\makeatletter

\DeclareRobustCommand{\greektext}{%
  \fontencoding{LGR}\selectfont\def\encodingdefault{LGR}}
\DeclareRobustCommand{\textgreek}[1]{\leavevmode{\greektext #1}}
\ProvideTextCommand{\~}{LGR}[1]{\char126#1}

\providecommand{\tabularnewline}{\\}

\usepackage{babel}
\usepackage{color}
\usepackage{dcolumn}
\usepackage{bm}
\usepackage{amsmath,amssymb}
\usepackage{graphicx}

\definecolor{background-color}{gray}{0.98}
\usepackage{hyperref}
\journal{Computer Physics Communications}

\@ifundefined{showcaptionsetup}{}{%
 \PassOptionsToPackage{caption=false}{subfig}}
\usepackage{subfig}
\makeatother

\usepackage{babel}
\begin{document}

\begin{frontmatter}{}

\title{Calculating Bardeen-Cooper-Schrieffer and Magnetic Superstructure Electronic States with \emph{\textgreek{JF}}}

\author{E.~Plekhanov$^{a,b,*}$, A.~Tchougréeff$^{b,c,d}$, R.~Dronskowski$^{c,e}$}

\address{$^{a}$King's College London, Theory and Simulation of Condensed
Matter (TSCM), The Strand, London WC2R 2LS, United Kingdom}

\address{$^{b}$Independent University of Moscow, Bol. Vlasevskiy per., 119002,
Moscow, Russia}

\address{$^{c}$Chair of Solid State and Quantum Chemistry, RWTH Aachen University,
52056 Aachen, Germany}

\address{$^{d}$A.N. Frumkin Institute of Physical Chemistry and Electrochemistry
of RAS, Moscow, Russia}

\address{$^{e}$ Jülich-Aachen Research Alliance, JARA-HPC, RWTH Aachen University,
52056 Aachen, Germany}

\cortext[cor]{Corresponding author. \emph{Email address: evgeny.plekhanov@kcl.ac.uk}}
\begin{abstract}
We propose the \emph{\textgreek{JF}} (Theta-Phi) package which addresses two of
the most important extensions of the essentially single-particle mean-field
paradigm of the computational solid state physics: the admission of
the Bardeen-Cooper-Schrieffer electronic ground state and allowance
of the magnetically ordered states with an arbitrary superstructure
(pitch) wave vector. Both features are implemented in the context
of multi-band systems which paves the way to an interplay with the
solid state \emph{quantum physics} packages eventually providing access
to the first-principles estimates of the relevant matrix elements
of the model Hamiltonians derived from the standard DFT calculations.
Several examples showing the workability of the proposed code are
given.
\end{abstract}

\end{frontmatter}{}

\section{INTRODUCTION\label{sec:Introduction} }

The solid state quantum physics packages available to the students
in the field are all based on the Hartree-Fock approximation for the
electronic wave function \citep{Anonymous2007-2017}. Extensions to
it are restricted to the so called ``post-Hartree-Fock'' methods
and largely reduce to the perturbation (Möller-Plesset order \emph{n}
- MP\emph{n}) corrections to the Hartree-Fock approximate ground state.
This significantly restricts the repertory of the types of the ground
states accessible to the available software. Practically, some of
the important types of the electronic states of solids cannot be reproduced
since they simply have not been programmed in. The most striking (and
scandalous) examples of unaccessible states are the Bardeen-Cooper-Schrieffer
(BCS) state \citep{Bardeen1957,Bardeen1957a} necessary for description
of superconductors. Even the intuitively more transparent states of
solids - the magnetically ordered ones with an arbitrary pitch (superstructure)
vector are not directly accessible by the available numerical tools.
The magnetically ordered states can be obtained by extending the chemical
unit cells to (magnetic) super-cells and setting the primeval magnetic
moments with \emph{broken symmetry} (BS) in the input file. Within
this technology only simplest ordered magnetic states - ferromagnetic
or antiferromagnetic with very simple magnetic super-lattices can
be accessed. Thus, the presence of either BCS or complex magnetic
states of solids is incurred indirectly. Namely, the presence of the
occupied non-bonding one-electron states in the vicinity of the Fermi
level indicates a possibility of that or another kind of instabilities
of the primary symmetric structure leading to some BS solution not
allowing to establish its specificity. Clearly, the BS solutions of
non-programmed types never come to surface. Another feature so far
missing in the available software is the temperature dependence of
the solutions of the electronic problems. This feature is, however,
important due to characteristic physical effects: transitions among
the high-temperature symmetric and various low-temperature BS phases
occurring in the experiment. Thus, we undertake the present development
with a goal to heel the outlined deficiencies in the existing software.
The paper is organized as follows: in Section \ref{sec:Theory-account}
we present necessary theoretical concepts; in Section \ref{sec:Implementation-details}
we describe implementation details; in Section \ref{sec:Examples}
we give test examples of applying developed software. In Section \ref{sec:Discussion-and-Conclusions}
we discuss the results of test calculations and give some perspectives.

\section{THEORY ACCOUNT\label{sec:Theory-account}}

\subsection{Hamiltonian}

The most general form of the Hamiltonian considered in $\Theta\Phi$
package is:
\begin{equation}
H=H_{K}+H_{U}+H_{J}+H_{V},\label{eq:Ham_gen}
\end{equation}
where $H_{K}$ is the one-electron part, containing the kinetic energy
and local terms:
\begin{equation}
H_{K}=\sum_{\substack{l,l^{\prime},R\\
\tau,s,s^{\prime}
}
}t_{l,l^{\prime}}^{s,s^{\prime}}(\tau)c_{l,R,s}^{\dagger}c_{l^{\prime},R+\tau,s^{\prime}}^{\phantom{\dagger}},\label{eq:H_K}
\end{equation}
$H_{U}$ is the on-site multi-orbital Coulomb repulsion:
\[
H_{U}=\sum_{\substack{l,l^{\prime},R\\
s,s^{\prime}
}
}U_{l,l^{\prime}}^{s,s^{\prime}}n_{l,R,s}n_{l^{\prime},R,s^{\prime}},
\]
and $H_{J}$ is the multi-site, multi-orbital Heisenberg term, which,
in general, can be anisotropic:
\[
H_{J}=\sum_{\substack{l,l^{\prime},R\\
\tau,\alpha,\beta
}
}J_{l,l^{\prime}}^{\alpha\beta}(\tau)S_{l,R}^{\alpha}S_{l^{\prime},R+\tau}^{\beta}.
\]
\label{eq:Spin_def}Finally, $H_{V}$ is the inter-site, inter-orbital
Coulomb repulsion:
\[
H_{V}=\sum_{l,l^{\prime},R,\tau}V_{l,l^{\prime}}(\tau)n_{l,R}n_{l^{\prime},R+\tau}.
\]
Here, $n_{l,R,s}=c_{l,R,s}^{\dagger}c_{l,R,s}^{\phantom{\dagger}}$
is the electron density operator on $l$-th orbital of site $R$,
while $n_{l,R}=n_{l,R,\uparrow}+n_{l,R,\downarrow}$ is the total
density operator and $S_{l,R}^{\alpha}$ is the operator for the spin
component $\alpha(=x,y,z)$:
\begin{align}
n_{l,R} & =\sum_{s}c_{l,R,s}^{\dagger}c_{l,R,s}^{\phantom{\dagger}}\nonumber \\
S_{l,R}^{\alpha} & =\frac{1}{2}\sum_{s,s^{\prime}}\sigma_{s,s^{\prime}}^{\alpha}c_{l,R,s}^{\dagger}c_{l,R,s^{\prime}}^{\phantom{\dagger}},\label{eq:Rho-components}
\end{align}
expressed as usual through Pauli matrices $\sigma_{s,s^{\prime}}^{\alpha}$.

In order to appropriately treat magnetic superstructure with arbitrary
orbital-dependent pitch vectors, we implement individual spin quantization
axis rotation for each orbital. Thus, there is no need to increase
the size of unit cell at the expense of modifying the Hamiltonian
terms. The orbital spin quantization axis rotation parameters together
with the superstructure pitch vector become additional variational
parameters and determine the optimal spin configurations. The details
of this technique are given in subsection\ref{subsec:Rotations}.

\subsection{Variational wave-function}

The Hamiltonian \eqref{eq:Ham_gen} is a true many-body operator,
impossible to solve exactly. We obtain an approximate variational
solution by optimizing its expectation value with respect to the variational
parameters of a trial wave-function within the mean-field solution.

The mean-field solution of eq.\eqref{eq:Ham_gen} is most easily obtained
with the use of the Nambu formalism: that is replacing the Fermi operators
$c_{l,R,s}^{\dagger}(c_{l,R,s}^{\phantom{\dagger}})$ creating an
electron (hole) with the spin projection $\sigma$ in the state single
state pertaining to the site (unit cell of the single band model)
by a Nambu vector 
\begin{equation}
\Psi_{R}^{\dagger}=\left(c_{1R\downarrow}^{\dagger},c_{1R\uparrow}^{\dagger},...,c_{LR\downarrow}^{\dagger},c_{LR\uparrow}^{\dagger},c_{1R\downarrow}^{\phantom{\dagger}},c_{1R\uparrow}^{\phantom{\dagger}},...,c_{LR\downarrow}^{\phantom{\dagger}},c_{LR\uparrow}^{\phantom{\dagger}}\right)\label{eq:Nambu}
\end{equation}
composed of the operators creating an electron (hole) with the spin
projection $s=\downarrow,\uparrow$ in one of the states $l=1,\dots,L$
in the unit cell $R$. Obviously, since $\Psi_{R}$ is a column, then
$\Psi_{R}^{\dagger}$ is a row consisting of corresponding hermitian
conjugated operators. Here we include in the basis both spin projections
as well as particle and hole creation operators in order to allow
for superconducting terms in the Hamiltonian as well as off-diagonal
spin exchange.

At this point we specify the class of the variational wave function
used in $\Theta\Phi$ which is a generalization of Anderson's RVB
wave function:

\begin{equation}
\left|\Psi_{g}\right\rangle =\prod_{\substack{j,k,\\
j<M
}
}\gamma_{j,k}^{\dagger}\left|\tilde{0}\right\rangle .\label{eq:Psi_RVB}
\end{equation}
Here $\gamma_{j,k}$ are the so-called canonical quasi-particles:
\[
\Gamma_{k}^{\dagger}=\left(\gamma_{1k}^{\dagger},\gamma_{2k}^{\dagger},\dots,\gamma_{4Lk}^{\dagger}\right).
\]
The number of operators in the Nambu vector $\Gamma_{k}$ is equal
to the length of $\Psi_{k}$: $4L$. There exists a unitary transformation
which relates the two:
\begin{equation}
\Gamma_{k}=\Xi_{k}\Psi_{k}.\label{eq:Xi_canonical}
\end{equation}
In eq.\eqref{eq:Psi_RVB} we have introduced a new vacuum $\left|\tilde{0}\right\rangle $
which is defined in a way, similar to that of the canonical quasiparticles'
vacuum in BCS theory\citep{Tinkham_1996}:
\[
\left|\tilde{0}\right\rangle =\prod_{j,k}c_{j,k,\downarrow}^{\dagger}\left|0\right\rangle .
\]
It is the coefficients of the matrices $\Xi_{k}$ which play the role
of the variational wave function parameters, although they are not
independent. In the present manuscript, we denote as $\left\{ \zeta\right\} $
the set of all \emph{independent} variables determining the $4L\times4L$
elements of matrices $\Xi_{k}$ in eq. \eqref{eq:Xi_canonical} at
all $k$. The term ``canonical'' for quasiparticles in the context
of $\left|\Psi_{g}\right\rangle $ means that this wave function is
build up by filling the vacuum $\left|\tilde{0}\right\rangle $ with
canonical quasi-particles. In other words, eq.\eqref{eq:Psi_RVB}
is a Slater determinant of canonical quasiparticles. In eq.\eqref{eq:Psi_RVB}
the filling occurs up to some band index $M$, which, in principle,
has to be considered as another variational parameter. In the case
of a particle-hole symmetric Hamiltonian, the optimal value of $M$
can be shown\citep{DeGennes_1999} to be $M=2L$ - a choice currently
realized in $\Theta\Phi$. The form of the wave function \eqref{eq:Psi_RVB}
as a set of independent quasiparticles uniquely defines the thermodynamics
of such an state: thermal excitations are quasiparticle-quasihole
pairs. This corresponds to the partition function being the product
of Fermi functions of quasiparticles energies determined in the self-consistent
procedure described below.

\subsection{Mean-field and self-consistency}

The mean-field solution of eq.\eqref{eq:Ham_gen} is accomplished
by minimizing the expectation value of either the variational energy
(at zero-temperature):
\begin{equation}
H_{var}=\frac{\left\langle \left.\Psi_{g}\right|H-\mu N\left|\Psi_{g}\right.\right\rangle }{\left\langle \Psi_{g}\left|\Psi_{g}\right.\right\rangle }\to\mathrm{min.}\label{eq:Evar}
\end{equation}
or of the Helmholtz free energy (at finite temperature) with respect
to a set of variational parameters $\left\{ \zeta\right\} $ (defined
below).
\begin{equation}
A=H_{var}-TS\to\mathrm{min}.\label{eq:Fvar}
\end{equation}
Here, $\mu$ is the chemical potential (taking care about the correct
number of electrons - see below), $T$ is the temperature in the energy
units (meaning $k_{B}=1$ in our notations), while $S$ is the (information/Shannon)
entropy (defined below). The number of particles operator is defined
as:
\[
N=\sum_{j,R}n_{j,R}.
\]
The necessary (but not sufficient) condition for the minimum of the(Helmholtz)
energy is the stationarity of the former with respect to the variational
parameters $\left\{ \zeta\right\} $ characterizing the trial wave
function eq.\eqref{eq:Psi_RVB}. The stationarity conditions can be
always expressed as a set of non-linear equations in the space of
variational parameters. This can be symbolically written as an equation:
\begin{equation}
\rho=\varPhi\left[\rho\right],\label{eq:Self-Consistency}
\end{equation}
where $\rho$ is the generalized one-particle density matrix of the
system corresponding to the wave function eq.\eqref{eq:Psi_RVB} and
$\varPhi$ symbolically represents the mean-field self-consistency
procedure allowing to derive the new density matrix given the old
one. The number of particles constraint narrows down even more the
search domain. In this restricted domain, the point giving the minimal
(Helmholtz) energy represents the solution to the problem eq.\eqref{eq:Evar}.
In addition, the equation $\left\langle N\right\rangle =N_{0}$ determines
the chemical potential $\mu$ and should be considered as a complement
to the system eq.\eqref{eq:Self-Consistency}.

Within the mean-field approach, the bare Hamiltonian $H$ is replaced
with a one-electron Hamiltonian $H^{MF}$:
\begin{align*}
H^{MF} & =H_{K}+H_{U}^{MF}+H_{J}^{MF}+H_{V}^{MF}.
\end{align*}
The terms with the superscript $MF$ are obtained from the respective
many-body operators by a ``linearization'' procedure (applying the
Wick's theorem\citep{Wick_1950}). The detailed forms of the linearized
operators are given in the \ref{sec:Appendix-A}. The fulfillment
of the condition eq.\eqref{eq:Fvar} amounts to solve a non-linear
optimization problem, which is a well-known challenge in both quantum
chemistry and solid state physics. It is outline in the following
subsections.

In terms of Nambu vectors the mean-field Hamiltonian reads: 
\[
H^{MF}=\sum_{R,\tau}\Psi^{\dagger}(R)\mathscr{H}(\tau)\Psi(R+\tau)+2L,
\]
where the last term accounts for the anti-commutation relations of
the operators comprising $\Psi_{R}$ ($L$ is the number of orbitals
of the problem). It brings a constant energy shift (arising from fermionic
anti-commutation rules) and will be omitted in what follows. The details
of the contributions to the Hamiltonian used are also given in the
\ref{sec:Appendix-A}. Here $\mathscr{H}(\tau)$ is the matrix representation
of $H^{MF}$ in the basis of operators, composing the Nambu vector
$\Psi(R)$.

Finally, we take advantage of periodicity of the crystal and transform
$H^{MF}$ to the reciprocal space:
\[
H^{MF}=\sum_{k}\Psi_{k}^{\dagger}\mathscr{H}_{k}^{\phantom{\dagger}}\Psi_{k}^{\phantom{\dagger}},
\]
with 
\[
\mathscr{H}_{k}=\sum_{\tau}e^{ik\tau}\mathscr{H}(\tau).
\]

\subsection{Density matrix and fixing $\mu$}

The central object in any mean-field theory is the density matrix,
which is defined through the Nambu notations eq.\eqref{eq:Nambu}
as:
\begin{equation}
\rho(\tau)=\mathbf{1}\delta_{\tau,0}-\langle\Psi_{R}^{\phantom{\dagger}}\Psi_{R+\tau}^{\dagger}\rangle.\label{eq:dens_mat_def}
\end{equation}
Here $\rho(\tau)$ are $4L\times4L$ matrices, independent on $R$
due to translational invariance and $\mathbf{1}$ is a $4L\times4L$
identity matrix. By using the hermiticity of$\mathscr{H}_{k}$ at
each $k$ we can define an arbitrary analytic function of matrix argument
in a usual way. As a consequence of the condition eq.\eqref{eq:Fvar},
the density matrix can be compactly written as:
\begin{equation}
\rho_{k}=\left(\mathbf{1}+e^{\beta\mathscr{H}_{k}}\right)^{-1},\label{eq:rho_def}
\end{equation}
where $\beta=1/T$ - inverse temperature. eq.\eqref{eq:rho_def} is
a generalization of Fermi-Dirac distribution. Here, we have also introduced
the Fourier transform of the density matrix:
\[
\rho_{k}=\sum_{\tau}e^{ik\tau}\rho(\tau).
\]

As described in Section \ref{sec:Implementation-details}, the density
matrix contains all the parameters necessary to define the trial wave
function eq.\eqref{eq:Psi_RVB} and the electronic phases of the system.
The optimization procedure is then equivalent to finding a self-consistent
solution for the density matrix obtained from eq.\eqref{eq:Self-Consistency}
\[
\rho_{new}=\varPhi[\rho_{old}],
\]
such that the newly generated $\rho_{new}$ does not deviate from
$\rho_{old}$ at the previous iteration within an accuracy $\eta_{\rho}$,
while the chemical potential is kept so that the total number of particles
is within an accuracy $\eta_{\mu}$ equal to $N_{0}$:
\begin{align}
\left\Vert \rho_{new}-\rho_{old}\right\Vert  & <\eta_{\rho}\nonumber \\
\label{eq:self-cons}\\
\left|N-N_{0}\right| & <\eta_{\mu}.\nonumber 
\end{align}

\subsection{\label{subsec:Rotations}Rotations of the local quantization axes}

In order to deal with arbitrary direction of spin-$\frac{1}{2}$ quantization
axes for each local orbital we implement the standard quantization
axis rotation formulae:
\begin{equation}
\Omega_{\mathbf{n},\vartheta}=\sigma_{0}\cos\frac{\vartheta}{2}-i\left(\mathbf{n},\boldsymbol{\mathbf{\sigma}}\right)\sin\frac{\vartheta}{2},\label{eq:SU2Matrix}
\end{equation}
which describes a rotation by angle $\vartheta$ around a rotation
axis with the unit vector $\mathbf{n}=(n_{x},n_{y},n_{z})$. Here
$\mathbf{\mathbf{\boldsymbol{\mathbf{\sigma}}}}=(\sigma_{x},\sigma_{y},\sigma_{z})$
is the vector composed of Pauli matrices, while $\sigma_{0}$ is the
$2\times2$ identity matrix.

In paper \citep{ArrigoniStrinati} it is shown that the Bloch states
for the electrons in a lattice can be built from the local states
defined with respect to the local (unit cell related) axes of the
spin quantization. These authors profit from the fact (explicitly
not formulated) that rotations in the spin space around the $y$-axis
form a commutative (sub)group (in fact the $SO(2)$ group) in the
general group $SU(2)$ of the spin rotations and thus its elements
can serve as images for a representation of the elements of another
commutative group: that of the lattice translations. The same is,
however, true for an arbitrary vector $\mathbf{n}$ fixed for all
unit cells of the lattice not necessarily for $\mathbf{n}=(0,1,0)$.
We use this option and allow the direction of $\mathbf{n}$ to be
an optimized variable. Anyhow the relative rotation angle between
two unit cells shifted by a lattice vector (one with integer components)
$\mathbf{\mathbf{\tau}}$ is given by 
\begin{equation}
\vartheta=\left(\mathbf{\tau},\mathbf{Q}\right)\label{eq:SuperstructureInducedAngle}
\end{equation}
where $\mathbf{Q}$ pitch or superstructure (wave)vector (belongs
to the reciprocal space). This defines the matrix
\begin{equation}
\Omega(\mathbf{n,\tau,Q})\label{eq:SuperstructureInducedSpinRotationMAtrix}
\end{equation}
 which must appear inside each product of the vector creation/destruction
operators representing the electron hopping between the unit cells
separated by the shift vector $\mathbf{\tau}$. For $\mathbf{\tau}=0$,
$\Omega(\mathbf{n,\tau,Q})=\sigma_{0}$.

It is not necessary, however, to assume that the spins within a unit
cell are all quantized along the same axis. That is to say that one
can assign to each orbital in the unit cell its own rotation matrix
determining the direction of its quantization axis in the global ``laboratory''
frame. Let $SU(2)$-matrices $\Omega(l)$ and $\Omega(l^{\prime})$
be those which rotate the spin quantization axes of the $l$-th and
$l^{\prime}$-th orbitals in the unit cell. Then for the pair of such
orbitals when the $l^{\prime}$-th orbital is located in the unit
cell shifted by the lattice vector $\mathbf{\tau}$ the matrix multiplier
\begin{equation}
\Omega^{\dagger}(l)\Omega(\mathbf{n,\tau,Q})\Omega(l^{\prime})\label{eq:GeneralSpinRotationMAtrix-2}
\end{equation}
must be inserted between the fermion-vector multipliers representing
the electron being destroyed in the $l^{\prime}$-th orbital and one
being created in the $l$-th orbital. This matrix is as well of the
form given by eq.\eqref{eq:SU2Matrix}. The complete formulae describing
the rotation of various Hamiltonian terms are reported in the \ref{sec:Rotations}.

\subsection{Physical properties}

Once the self-consistency is reached, the observables: correlation
functions as well as band structure and electronic density of states
(DOS) can be calculated. Below we summarize the formulas for some
of them, but many others can be defined in addition. At the self-consistency,
the internal energy $U$ is just an average of the mean field Hamiltonian:
\[
U=\left\langle H^{MF}\right\rangle =\frac{\sum_{k}\mathrm{Tr}\left(\mathscr{H}_{k}\rho_{k}\right)}{\sum_{k}\mathrm{Tr}\rho_{k}}.
\]
The (Helmholtz) free energy $A$ and the entropy $\varsigma$ are
defined through the spectrum of the canonical quasi-particles:
\begin{equation}
A=-T\sum_{k}\mathrm{Tr}\ln\left(\mathbf{1}+e^{-\beta\mathscr{H}_{k}}\right),\label{eq:F_quasi}
\end{equation}
\[
S=-\frac{\partial A}{\partial T}=\frac{U-A}{T}.
\]
The expression eq.\eqref{eq:F_quasi} is used in practical calculations
as anticipated above. The specific heat can also be easily obtained:
\[
C_{V}=\frac{\partial U}{\partial T}=-T\frac{\partial^{2}A}{\partial T^{2}}.
\]
The uniform Pauli magnetic susceptibility is given by:
\[
\chi_{P}=\frac{\partial M_{tot}^{z}}{\partial\mathcal{H}_{z}},
\]
where $M_{tot}^{z}$ is the total magnetic moment of the site (unit
cell):
\[
M_{tot}^{z}=\sum_{l}M_{0,l}^{z}=\frac{1}{2}\sum_{l}\left\{ \left\langle c_{l0\uparrow}^{\dagger}c_{l0\uparrow}\right\rangle -\left\langle c_{l0\downarrow}^{\dagger}c_{l0\downarrow}\right\rangle \right\} ,
\]
expressed through the density matrix elements between the local states.

The band structure along the path connecting the high symmetry points
of the Brillouin zone as well as the DOS are both obtained from the
spectral density $A_{l}(k,\epsilon)$:
\[
A_{l}(k,\epsilon)=-\frac{1}{\pi}\mathrm{Im}\left[\left(\left(\epsilon+i\delta\right)\mathbf{1}-\mathscr{H}_{k}\right)^{-1}\right]_{ll},
\]
so that the total DOS: $\sum_{k,l}A_{l}(k,\epsilon)$, the partial
one: $\sum_{k}A_{l}(k,\epsilon)$ and the band structure (as a set
of points in $\epsilon,k$ plane where $\sum_{l}A_{l}(k,\epsilon)$
diverges) can be obtained.

Closing the discussion of the observables, we note that the solutions
of self-consistency equations have their domains of existence as functions
of Hamiltonian parameters, filling, temperature and any other external
condition. These domains form the system's phase diagrams, which can
be directly compared with the experiment and/or other theories etc.
As well, there might exist (and this is the standard situation) several
solutions or phases for the same values of the Hamiltonian parameters,
and/or temperature etc. In this case, owing to the variational principle,
the phase with the lowest free energy must be chosen, which paves
the way to the phase transitions at finite temperature. 

\section{IMPLEMENTATION DETAILS\label{sec:Implementation-details}}

The general energy optimization procedure as applied to the electronic
phases of a crystal can be in principle solved in two ways:
\begin{enumerate}
\item as a self-consistency procedure: starting as described above from
a trial $\rho_{old}$ and iteratively calculating new $\rho_{new}$
at each step finding the chemical potential. In order to stabilize
such a procedure usually some forms of mixing are used. Here we use
the simplest linear mixing, such that $\rho_{new}=\alpha\rho_{new}+(1-\alpha)\rho_{old}$
with $0\leqslant\alpha\leqslant1$;
\item the self-consistency condition eq.\eqref{eq:Self-Consistency} supplemented
with the constraint fixing the number of particles can be also solved
by using the globally convergent Newton-Raphson method\citep{NumRec}
for root finding. It can be shown, that the self-consistency condition
is equivalent to considering the evanescence conditions for the derivatives
of the energy with respect to the matrix elements of the density.
Finally, the energy is compared at each step to ensure the the chosen
trial step actually leads to a minimum.
\end{enumerate}
Both approaches are implemented in $\Theta\Phi$. In the second case
it is crucial to use all possible symmetries of $\rho$ in order to
reduce the number of unknowns. This will be highlighted in the following
subsection. As will become evident from the following subsection,
the density matrix contains all the (order) parameters necessary to
define the trial wave function eq.\eqref{eq:Psi_RVB}.

\subsection{Broken symmetry and generalized density matrix\label{subsec:Broken-symmetry}}

In this subsection we briefly discuss the properties of the generalized
density matrix $\rho$. This is the central object of mean-field theory
which determines the electronic phase of the system. Variational wave
function \eqref{eq:Psi_RVB} depends on a set of variational parameters
$\left\{ \zeta\right\} $. On the other hand, the density matrix $\rho$,
defined as a matrix of pairwise averages of all fermionic operators
present in the Nambu vector eq.\eqref{eq:Nambu}, also depends on
$\left\{ \zeta\right\} $. The number of independent parameters $\left\{ \zeta\right\} $
is in general much smaller than the number of elements in $\rho$.
However, analyzing $\rho$ helps to give a physical meaning to the
variational parameters $\left\{ \zeta\right\} $ as well as to determine
the electronic phase of the system. Moreover, by triggering some of
$\left\{ \zeta\right\} $ it is possible to induce a phase change
in the system. The wave function eq.\eqref{eq:Psi_RVB} allows for
two solutions breaking fundamental symmetries (and their combination):
i) particle number conservation and ii) time-reversal invariance.
The former corresponds to the superconducting (BCS) states, while
the later - to all kinds of magnetic states (when supplied with the
spin quantization axis rotation). In this context, we call the inter-relations
among the $\rho$'s matrix elements (or constraints on their values)
- \emph{a particular density matrix symmetr}y. These symmetries should
not be confused with the spatial or point group ones. Typically, the
so-called anomalous terms in $\rho$ (those of the form $\left\langle c_{\uparrow}^{\dagger}c_{\downarrow}^{\dagger}\right\rangle $
or $\left\langle c_{\downarrow}^{\phantom{\dagger}}c_{\uparrow}^{\phantom{\dagger}}\right\rangle $)
if non-vanishing lead to BCS states, while the imbalance between ``up''
and ``down'' spin averages ($\left\langle c_{\uparrow}^{\dagger}c_{\uparrow}^{\phantom{\dagger}}\right\rangle \neq\left\langle c_{\downarrow}^{\dagger}c_{\downarrow}^{\phantom{\dagger}}\right\rangle $)
leads to magnetically ordered phases. Finally, we note that $\rho(\tau=0)$
is in general Hermitian, while at a finite lattice shifts this is
not true any more and the following relation holds:

\begin{eqnarray*}
\rho(\tau) & = & \rho^{\dagger}(-\tau),
\end{eqnarray*}
\emph{i.e.} Hermitian conjugate of a density matrix at the lattice
shift $\tau$ is the density matrix at the lattice shift $-\tau$
assuring the hermiticity of the density matrix as a whole.

To be more specific, we adapt the general definition eq.\eqref{eq:dens_mat_def}
to the case of a two-orbital system where we allow for all possible
types of pairing. The density matrix at zero distance in that case
can be written as:

\[
\rho(\tau=0)=\begin{pmatrix}n_{1\downarrow} & 0 & r_{12\downarrow} & 0 & 0 & -\Delta_{1}^{\star} & 0 & -\Delta_{12}^{\star}\\
0 & n_{1\uparrow} & 0 & r_{12\uparrow} & \Delta_{1}^{\star} & 0 & \Delta_{12}^{\star} & 0\\
r_{12\downarrow} & 0 & n_{2\downarrow} & 0 & 0 & -\Delta_{12}^{\star} & 0 & -\Delta_{2}^{\star}\\
0 & r_{12\downarrow} & 0 & n_{2\uparrow} & \Delta_{12}^{\star} & 0 & \Delta_{2}^{\star} & 0\\
0 & \Delta_{1} & 0 & \Delta_{12} & 1-n_{1\downarrow} & 0 & -r_{12\downarrow} & 0\\
-\Delta_{1} & 0 & -\Delta_{12} & 0 & 0 & 1-n_{1\uparrow} & 0 & -r_{12\uparrow}\\
0 & \Delta_{12} & 0 & \Delta_{2} & -r_{12\downarrow} & 0 & 1-n_{2\downarrow} & 0\\
-\Delta_{12} & 0 & -\Delta_{2} & 0 & 0 & -r_{12\uparrow} & 0 & 1-n_{2\uparrow}
\end{pmatrix}.
\]
Here the parameters $n_{1(2)\sigma},\Delta_{1(2)},\Delta_{12},r_{12\sigma}$
have the meaning of local occupation, local and inter-orbital pairing
and inter-orbital bonding respectively. The density matrix at $\tau\ne0$
has the same structure, except for the fact that on the diagonal there
would be the inter-site intra-orbital hopping: $n_{1(2)\sigma}\to h_{1(2)\sigma}$
and all the other parameters will be sampled at the corresponding
distance $r$. The zero matrix elements of $\rho$ stem from the definition
of the trial wave-function \eqref{eq:Psi_RVB} and from the commutation
relations for fermionic operators. Already at this step it is clear
that a $16\times16$ matrix actually has only $9$ independent matrix
elements. This symmetry is checked and enforced during $\Theta\Phi$
run, using iterative self-consistent scheme, while the reduction from
$\rho(\tau=0)$ to parameters $n_{1(2)\sigma},\Delta_{1(2)},\Delta_{12},r_{12\sigma}$
and back is used in Newton-Raphson method in order to reduce the number
of self-consistency equations and unknowns.

The switching between different phases in eq.\eqref{eq:Psi_RVB} is
accomplished by zeroing some of the parameters. \emph{E.g.} setting
all $\Delta=0$ brings the system into a Fermi-liquid phase with no
superconducting fluctuations. Setting $\Delta_{1(2)}=0$ excludes
local pairing, setting $\Delta_{12}$ on the contrary forbids the
inter-orbital pairing. The well-known $d_{x^{2}-y^{2}}$-wave pairing
can be achieved on a square lattice by imposing that the superconducting
part of the density matrix along \emph{e.g.} $y$ direction be equal
to minus that along $x$ direction, which in case of a single orbital
model can be symbolically written as: 

\[
\rho_{c_{\uparrow}^{\dagger},c_{\downarrow}^{\dagger}}(r\parallel x)=-\rho_{c_{\uparrow}^{\dagger},c_{\downarrow}^{\dagger}}(r\parallel y).
\]
Of course, in case of multi-orbital systems and complicated lattices
even more involved inter-relations between $\rho$ matrix elements
at various distances (see e.g. Ref.\citep{Black2014}), may exist
which are advisable to be analyzed before using $\Theta\Phi$.

\subsection{Matrix elements and porting with existing programs}

Thus, $\Theta\Phi$ requires the hoping matrix elements $t_{l,l^{\prime}}^{s,s^{\prime}}(\tau)$
as well as $U_{l,l^{\prime}}^{s,s^{\prime}}$,$V_{l,l^{\prime}}(\tau)$,
and $J_{l,l^{\prime}}^{\alpha\beta}(\tau)$, as input. While the user
can specify all the Hamiltonian parameters manually, assuming an abstract
theoretical model, we additionally introduced a functionality allowing
the import of the hopping matrix elements $t_{l,l^{\prime}}^{s,s^{\prime}}(\tau)$
from existing first-principles codes. This is achieved by reading
the tight-binding Hamiltonians, generated by either of the two well-known
post-processing codes: \textsc{wannier90} \citep{wannier90} and \textsc{lobster}
\citep{Tch095,Lobster_1,Lobster_2}. By doing this, the precise, material-specific
predictions can be made. $\Theta\Phi$ will read the files called
wannier90\_hr.dat (for \textsc{wannier90}) or RealSpaceHamiltonians.lobster
(for \textsc{lobster}) and populate the hopping matrix, superseding
any previously stored. Both \textsc{wannier90} and \textsc{lobster}
have in turn interfaces to all major \emph{ab-initio} codes (like
\textsc{vasp}, \textsc{abinit}, \textsc{castep}, \textsc{QuantumEspresso}
etc.), which allows the $\Theta\Phi$ user to profit from the results
obtained by all these codes. Additionally, we can recommend using
the \textsc{MagAîxTic} package \citep{Tch105} to obtain the required
estimates of the$J_{l,l^{\prime}}^{\alpha\beta}(\tau)$ parameters.

\section{EXAMPLES\label{sec:Examples}}

The present examples Section pursues a dual purpose: to illustrate
the capacities of the proposed package and by this to contribute to
bridging certain gap between solid state quantum chemistry and solid
state theoretical physics communities. For the former, the variety
of the electronic states/phases considered and studied by the latter
remains to a significant extent inaccessible together with the fascinating
properties of these phases since they are simply not present in the
tools (solid state quantum chemistry codes) the former use. For the
latter, incidentally, remains inaccessible the possibility to independently
estimate the parameters (matrix elements) of the model Hamiltonians
featuring the most interesting phases/hottest topics of experimental
and theoretical interest. Thus in this Section we start with simple
example of the 1D Hubbard model.

\subsection{Superconducting state in 1D Hubbard model}

We start by considering the simplest model for strongly correlated
electrons on a 1D ring of sites, each holding one orbital - single-orbital
Hubbard model, which has the Hamiltonian:

\begin{equation}
H=-t\sum_{r,s}\left(c_{rs}^{\dagger}c_{r+1s}^{\phantom{\dagger}}+h.c.\right)+U\sum_{r}n_{r\uparrow}n_{r\downarrow}.\label{eq:Hubbard_ham}
\end{equation}
Here, we denote by $r$ the coordinate of the current cell (site),
and we treat the interaction term according to the prescriptions outlined
in Appendix \ref{sec:Appendix-A}. If $U<0$, it is possible to stabilize
a simple BCS superconducting solution, thanks to the electron-electron
attraction term in the BCS Hamiltonian: $U\langle c_{r\uparrow}^{\dagger}c_{r\downarrow}^{\dagger}\rangle\langle c_{r\downarrow}^{\phantom{\dagger}}c_{r\uparrow}^{\phantom{\dagger}}\rangle$.
The BCS state corresponds to the density matrix of the form:

\[
\rho=\begin{pmatrix}n/2 & 0 & 0 & -\Delta^{\star}\\
0 & n/2 & \Delta^{\star} & 0\\
0 & \Delta & 1-n/2 & 0\\
-\Delta & 0 & 0 & 1-n/2
\end{pmatrix}.
\]
Here the components of the density matrix are: $n=\left\langle c_{rs}^{\dagger}c_{rs}^{\phantom{\dagger}}\right\rangle $,
$\Delta=\langle c_{r\uparrow}^{\dagger}c_{r\downarrow}^{\dagger}\rangle$
being respectively the site population and the anomalous average.
In the case of an on-site Hubbard attraction, only $\rho$ at zero
lattice shift enters into the self-consistency equations, and the
above averages do not depend on $r$ due to the translational invariance.

\begin{figure}
\includegraphics[angle=270,width=0.5\columnwidth]{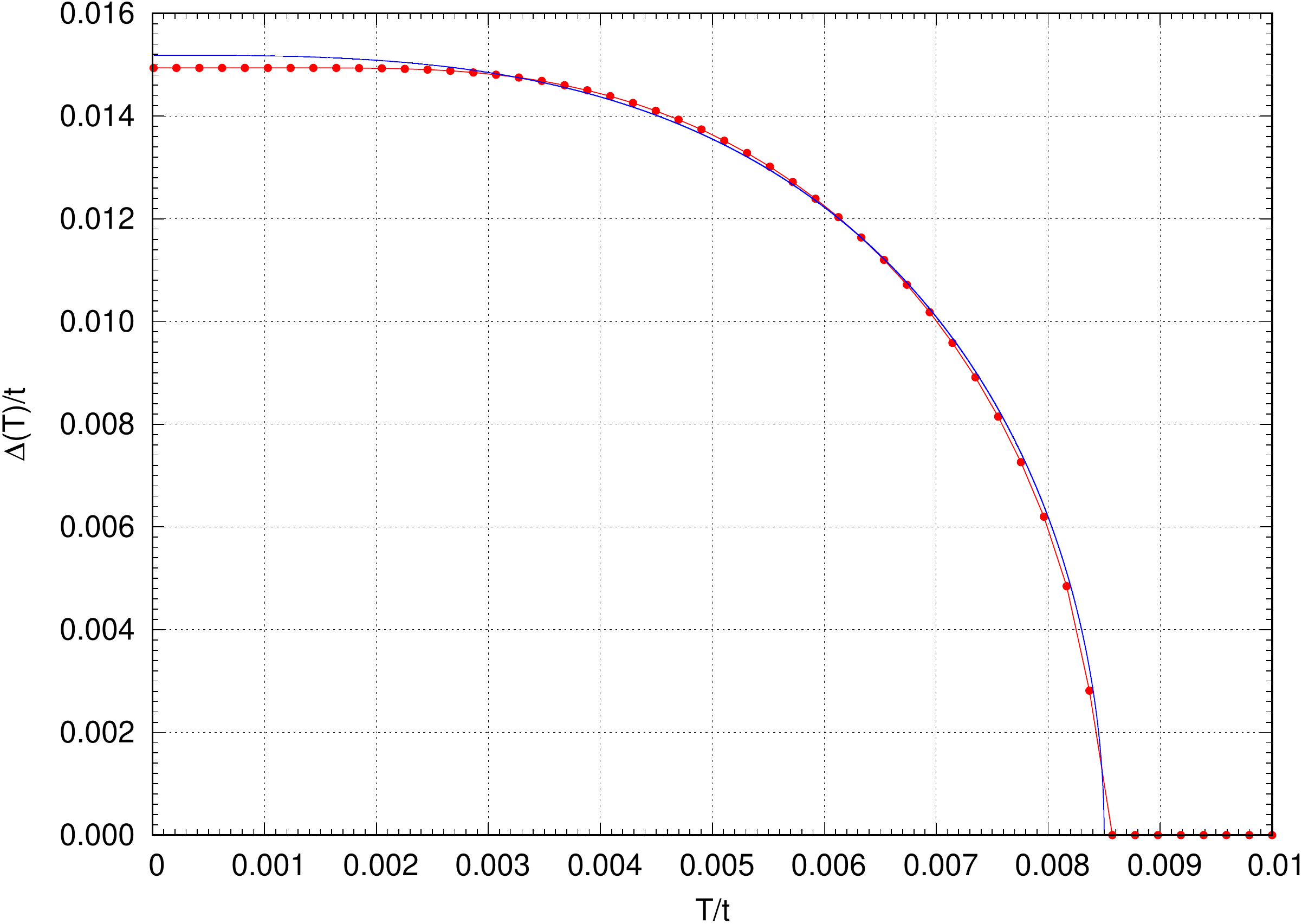}\includegraphics[angle=270,width=0.5\columnwidth]{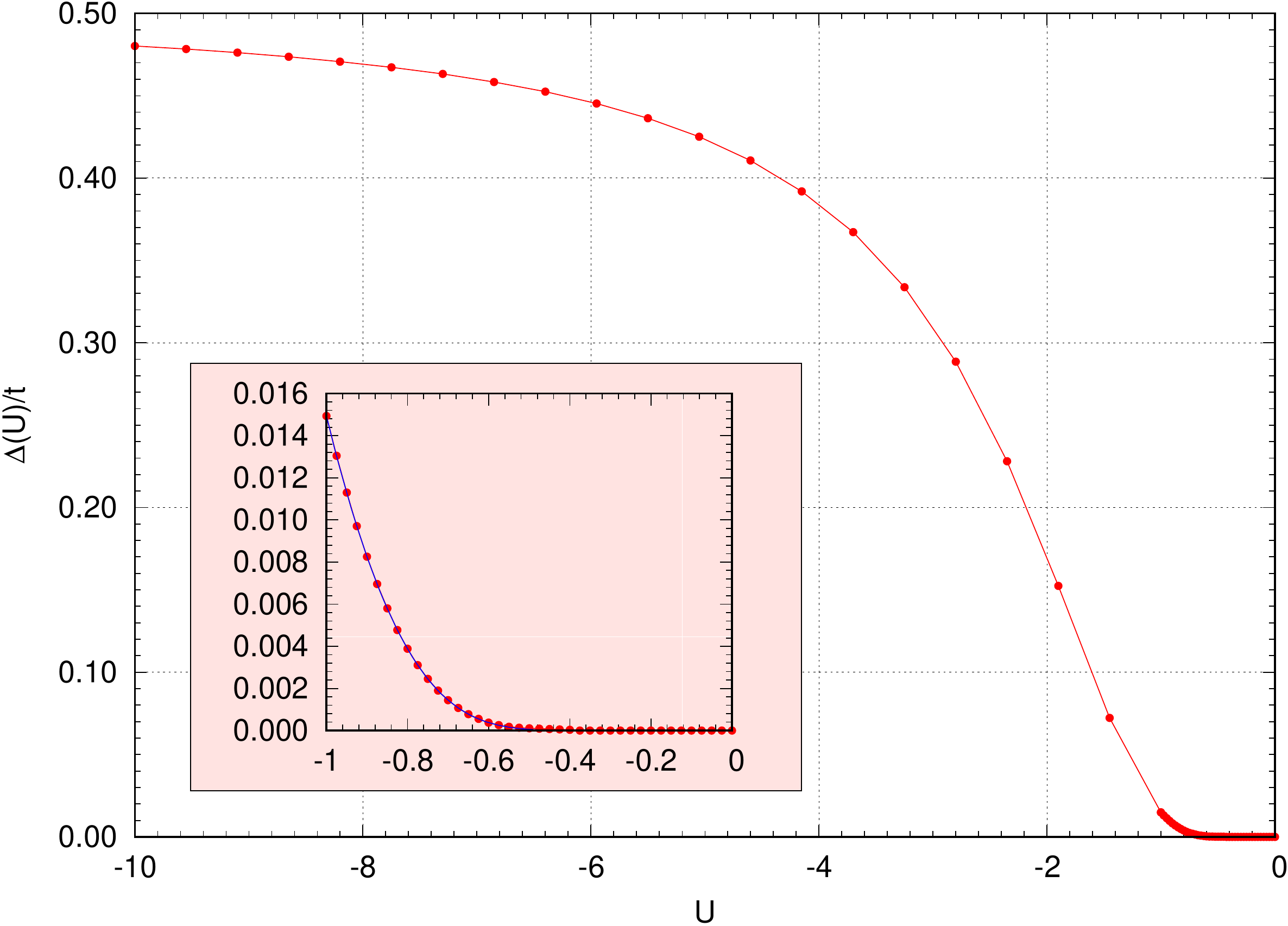}

\caption{\label{fig:Hubbard}$U=-t$, temperature (left panel) and Hubbard-$U$
(right panel) dependence at $T=10^{-5}t$ of the superconducting paring
$\Delta$ (corresponding to $\rho_{14}(\tau=0)$) for local superconducting
phase of 1D Hubbard model. $n=1$ (half-filling) $50000$ sites. The
inset shows the details of how the superconducting element of the
density ($\Delta$) approaches zero in $U\to0$ limit. The blue lines
inset are the respective analytic formulae \eqref{eq:D_vs_T} and
\eqref{eq:D_vs_U}.}
\end{figure}

The $\Theta\Phi$ results for this simple model eq.\eqref{eq:Hubbard_ham}
are shown in Fig.\eqref{fig:Hubbard}. As expected from the BCS theory,
the temperature dependence of the a superconducting order parameter
$\Delta$is described with a very good accuracy by: 
\begin{equation}
\Delta(T)=\Delta_{0}\sqrt{1-\left(\frac{T_{\phantom{c}}}{T_{c}}\right)^{3}},\label{eq:D_vs_T}
\end{equation}
in the vicinity of the transition temperature $T_{c}$. This is very
remarkable feature provided by the \textsc{\ensuremath{\mathit{\Theta}}\ensuremath{\mathit{\Phi}}}
program - the temperature dependence of the solution of the electronic
problem. Incidentally, the $U$-dependence of $\Delta_{0}$ at lowest
temperature is very well described by a BCS-like formula:
\begin{equation}
\Delta_{0}(U)=-\frac{8e^{\frac{2\pi}{U}}}{U},\label{eq:D_vs_U}
\end{equation}
for small absolute values of \emph{negative i.e. }attractive $U$.\footnote{We note that this simple variational ground state does not represent
adequately the \emph{true} ground state of the 1D Hubbard model. Indeed,
by using the bosonization technique\citep{bosonization_2004}, it
was demonstrated that in a wide range of 1D models with short-ranged
interactions there occurs the separation of spin and charge degrees
of freedom, while the correlation functions decay with a power-law
as a function of distance. Thus the more accurate treatment reveals
no long range correlations in 1D systems, although a sizable superconducting
correlations are also predicted by bosonization.} By this we demonstrate the workability of two important features
of $\Theta\Phi$ - the accessibility of the BCS state of the system
and the temperature dependence of the solution of an electronic problem.

\subsection{Superconducting states of Graphene\label{subsec:RVB-states-of}}

Superconductivity in pristine graphene does not occur naturally, although
recently the superconductivity has been reported in graphene bilayers
rotated at a series of ``magic'' angles\citep{cao_correlated_2018,cao_unconventional_2018}.
On the other hand, combination of Dirac-like linear spectrum with
superconductivity may lead to a bunch of novel phenomena, such as
specular Andreev reflection \citep{Beenakker_2006}. Moreover, recently
an artificial honeycomb 2D super-lattice - effectively equivalent
to graphene layer - has been realized by using the self-assembly of
PbSe, PbS or CdSe nano-crystals matching their (100) facets \citep{Boneschanscher_2014}.
Such super-lattices, possess well defined Dirac spectrum, although
with several new features arising from stronger spin-orbit coupling
\citep{Kalesaki_2014,beugeling_2015}. Several approaches predicted
superconductivity with various order parameter symmetries on 2D honeycomb
lattice, among which auxiliary-boson theory (ABF) \citep{Black2007,Black2014}
(also called the Resonating Valence Bond - RVB - state in this context),
renormalization-group\citep{Nandkishore_2012}, or even self-consistent
Bogoliubov-de Gennes approach to spatially inhomogeneous superconductivity
\citep{Potirniche_2014,Nandkishore_2012}.

\begin{figure}
\includegraphics[viewport=-70bp -60bp 297bp 294bp,width=0.33\textwidth]{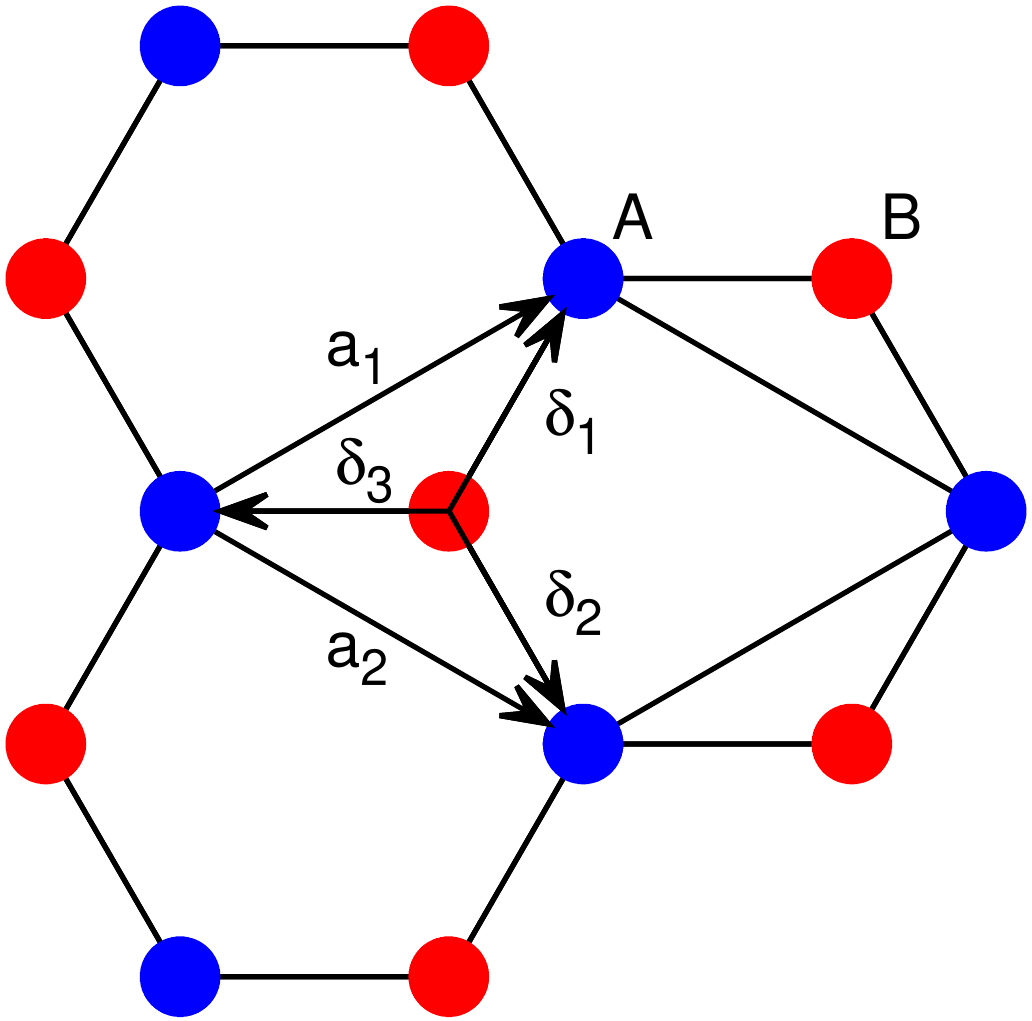}\hfill{}\includegraphics[viewport=0bp 0bp 622bp 423bp,width=0.5\columnwidth]{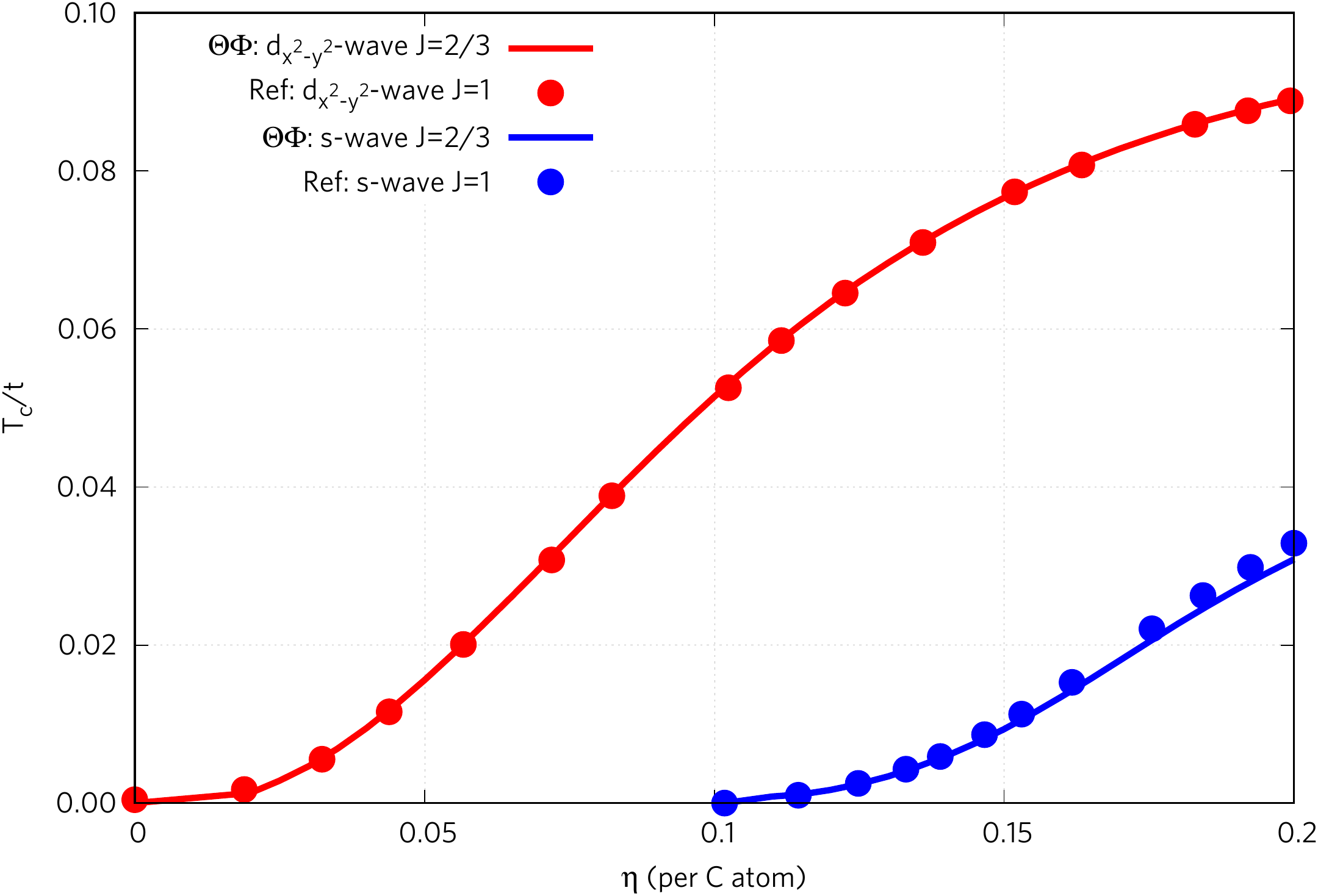}\caption{\label{fig:Graphene-unit-cell}Left panel: graphene unit cell showing
two types of sites ($A$ and $B$), vectors $\{\delta_{i}\}$ connecting
NN sites, and the unit cell vectors $\{a_{1(2)}\}$. Picture taken
from Ref. \citep{RevCastro2009}. Right panel: Comparison of the doping
dependence $\theta_{c}(\eta)$ for $s$-wave and $d_{x^{2}-y^{2}}$-wave
phases of graphene with the data from Ref.\citep{Black2014}. The
points were extracted from the graphs of Ref. \citep{Black2014},
while the lines were calculated by $\Theta\Phi$. For the details
of the calculations and the difference in the exchange-$J$ definitions
see in the text.}
\end{figure}

Here, we benchmark $\Theta\Phi$ by reproducing several ABF superconducting
solutions of various symmetries in graphene and by comparing them
with the reference results \citep{Black2007,Black2014}. We remind
that on 2D honeycomb lattice each lattice site connects to three nearest
neighbor (NN) sites by vectors $\{\delta_{i}\},\;i=1,2,3$. These
vectors can be defined in several ways, and here we adopt the following
definition (see Fig. \ref{fig:Graphene-unit-cell}):
\[
\delta_{1}=a\left(\begin{array}{c}
\frac{1}{2}\\
\frac{\sqrt{3}}{2}
\end{array}\right),\quad\delta_{2}=a\left(\begin{array}{c}
\phantom{-}\frac{1}{2}\\
-\frac{\sqrt{3}}{2}
\end{array}\right),\quad\delta_{3}=a\left(\begin{array}{c}
-1\\
\phantom{-}0
\end{array}\right).
\]
The unit cell of graphene contains two atoms and is defined by the
following unit cell vectors: $a_{1}=\delta_{1}-\delta_{3}$ and $a_{2}=\delta_{2}-\delta_{3}$.
If the coordinates of the first site in the unit cell are chosen to
be zero, then the coordinates of the second one will be $u=\delta_{1}+\delta_{2}-\delta_{3}$.
With only NN hopping active, each unit cell is connected to its four
nearest neighbors at lattice shifts $\tau=\pm a_{1},\pm a_{2}$.

Superconducting phases on honeycomb lattice can come out with several
symmetries: extended $s$-wave, $d_{x^{2}-y^{2}}$-wave, $d_{xy}$-wave
and several triplet pairing symmetries (like $p$-wave); corresponding
to different fixed relations between the superconducting order parameters
$\langle c_{r\uparrow}^{\dagger}c_{r+\delta_{i}\downarrow}^{\dagger}\rangle$.
In the present work, we consider only the extended $s$-wave, $d_{x^{2}-y^{2}}$-wave,
$d_{xy}$-wave and report the $k$-space dependence of their gap functions
in Tab. \ref{tab:tab_symm}.

\begin{table}
\begin{tabular}{|c|c|c|c|}
\hline 
 & extended-$s$ & $d_{x^{2}-y^{2}}$ & $d_{xy}$\tabularnewline
\hline 
\hline 
 &  &  & \tabularnewline
$\Delta(\mathbf{k})\sim$ & $e^{i\mathbf{k}\delta_{1}}+e^{i\mathbf{k}\delta_{2}}+e^{i\mathbf{k}\delta_{3}}$ & $e^{i\mathbf{k}\delta_{2}}-e^{i\mathbf{k}\delta_{3}}$ & $2e^{i\mathbf{k}\delta_{1}}-e^{i\mathbf{k}\delta_{2}}-e^{i\mathbf{k}\delta_{3}}$\tabularnewline
\hline 
\end{tabular}

\caption{\label{tab:tab_symm}$k$-dependence of the gap function in case of
extended-$s$, $d_{x^{2}-y^{2}}$, and $d_{xy}$-wave superconductivity
on 2D honeycomb lattice. The vectors $\{\delta_{i}\}$, connecting
a graphene site to its three nearest neighbors are defined in the
text.}

\end{table}

The Hamiltonian of free electrons, hopping at NN distance on honeycomb
lattice, reads as follows:

\[
H_{0}=-t\sum_{\delta,r}c_{rs}^{\dagger}c_{r+\delta s}^{\phantom{\dagger}},
\]
where $t$ is the hopping matrix element. It can be readily diagonalized,
yielding the well-known graphene spectrum:

\[
E(k_{x},k_{y})=\pm t\sqrt{1+4\cos{\frac{\sqrt{3}k_{x}}{2}}\cos{\frac{k_{y}}{2}}+4\cos^{2}{\frac{k_{y}}{2}}}.
\]
It manifests the Dirac cones at $K=\pi(\frac{1}{3},\frac{2}{3})$
and $K^{\prime}=\pi(\frac{2}{3},\frac{1}{3})$ points of the Brillouin
zone.

In Ref. \citep{Black2007}, it was argued that a sizable exchange
correlation $J$ in graphene could arise from on-site Hubbard repulsion
$U$: $J=\frac{2t^{2}}{U}.$ We remind the value of $t$ known for
long time in graphene: $t=2.8$ \foreignlanguage{english}{eV}, (see
\citep{Tch021,Tch039} and reference therein) while $U$ estimated
in Ref. \citep{Black2007,Black2014} is $U=3.3t$. These estimates
give for the ratio $J/t\approx1.7$.

With the exchange term the full interacting Hamiltonian becomes:
\begin{equation}
H_{gra}=H_{0}+J\sum_{\delta,r}S_{r}S_{r+\delta},\label{eq:H_gra}
\end{equation}
where the spin operators are defined through the fermionic operators
according to eq.\eqref{eq:Rho-components}.

We treat the Hamiltonian eq.\eqref{eq:H_gra} within the mean-field
approach in $\Theta\Phi$ according to the formalism outlined in the
previous Section\ref{sec:Theory-account} and compare our results
with those of Refs. \citep{Black2007,Black2014}. In the case of graphene,
which has two atoms in the unit cell (site $A$ and site $B$), we
set $L=2$. Therefore, the density matrix in real space is a $8\times8$
matrix for each distance. Here we consider the $t-J$ model on honeycomb
lattice in which all the terms (both kinetic and interaction parts)
have the range extending to at most NN inter-cell lattice shift $\tau=a_{1(2)}$,
as shown in Fig.\ref{fig:Graphene-unit-cell}. In addition, there
is always present the local density matrix at $\tau=0$. For the sake
of illustration, we show below local part of $\rho$:

\begin{equation}
\rho(\tau=0)=\begin{pmatrix}n_{1\downarrow} & 0 & r_{12\downarrow} & 0 & 0 & 0 & 0 & -\Delta_{12}^{\star}\\
0 & n_{1\uparrow} & 0 & r_{12\uparrow} & 0 & 0 & \Delta_{12}^{\star} & 0\\
r_{12\downarrow} & 0 & n_{2\downarrow} & 0 & 0 & -\Delta_{12}^{\star} & 0 & 0\\
0 & r_{12\downarrow} & 0 & n_{2\uparrow} & \Delta_{12}^{\star} & 0 & 0 & 0\\
0 & 0 & 0 & \Delta_{12} & 1-n_{1\downarrow} & 0 & -r_{12\downarrow} & 0\\
0 & 0 & -\Delta_{12} & 0 & 0 & 1-n_{1\uparrow} & 0 & -r_{12\uparrow}\\
0 & \Delta_{12} & 0 & 0 & -r_{12\downarrow} & 0 & 1-n_{2\downarrow} & 0\\
-\Delta_{12} & 0 & 0 & 0 & 0 & -r_{12\uparrow} & 0 & 1-n_{2\uparrow}
\end{pmatrix},\label{eq:Rho_Graph}
\end{equation}
where various parameters appearing inside eq.\eqref{eq:Rho_Graph}
are $\tau=0$ limit of the following correlation functions:

\begin{alignat*}{2}
n_{ls}(\tau) & =\langle c_{lRs}^{\dagger}c_{lR+\tau s}^{\phantom{\dagger}}\rangle & \textrm{ intra-sublattice hopping}\phantom{.}\\
r_{12s}(\tau) & =\langle c_{1Rs}^{\dagger}c_{2R+\tau s}^{\phantom{\dagger}}\rangle & \textrm{inter-sublattice hopping\ensuremath{\phantom{.}}}\\
\Delta_{12}(\tau) & =\langle c_{1R\uparrow}^{\dagger}c_{2R+\tau\downarrow}^{\dagger}\rangle & \textrm{ inter-sublattice sc. paring.}
\end{alignat*}
Here, $i=1,2$ is the orbital index, $\tau$ is the given unit cell
separation, $s$ is the spin index. In addition, we consider only
the non-local superconducting averages.

In the case of the extended-$s$ phase, all $\Delta_{12}$ at lattice
shifts $\tau=0$, $\tau=a_{1}$ and $\tau=a_{2}$ are equal, while
in the $d_{x^{2}-y^{2}}$ case: $\Delta_{12}=0$ at $\tau=0$ and
$\Delta_{12}(\tau=a_{1})=-\Delta_{12}(\tau=a_{2})$, finally in the
$d_{xy}$ case: $\Delta_{12}(\tau=0)=-2\Delta_{12}(\tau=a_{1})=-2\Delta_{12}(\tau=a_{2})$.
$\sum_{i,\sigma}n_{i\sigma}(\tau=0)=\mathscr{\mathcal{N}}$ fixes
the chemical potential by imposing the average occupation per unit
cell to be $\mathcal{N}=2$. The angular dependence of the superconducting
gap in $k$-space for the three symmetry cases\citep{Black2014} is
summarized in Table\ref{tab:tab_symm}.

A few words should be said about the treatment of the exchange term
in $\Theta\Phi$ and in the reference articles. Namely, in the former
we retain all non-superconducting terms in the mean-field decoupling
of the Hamiltonian term proportional to $J$ (see \ref{sec:Appendix-A}),
while in the latter the hopping renormalization, arising from the
linearized exchange term was not taken into account. That is to say,
discarding the last term in eq.\eqref{eq:HJ}. Moreover, when making
the mean-field average, the coefficient of $\frac{3}{2}$ in front
of $J$ was omitted too. In order to compare our results with those
of Ref. \citep{Black2014}, we rescale $J$ down to a factor of $\frac{2}{3}$
and discard the hopping renormalization in this subsection. With this
assumptions, the agreement with the reference data of Ref. \citep{Black2014}
is excellent, as can be seen from the right panel of Fig.\ref{fig:Graphene-unit-cell},
where we compare the critical superconducting temperature $T_{c}$
as a function of doping $\eta$ (per C atom) at $J=1$ in the notations
of Ref. \citep{Black2014}. This agreement holds for both extended-$s$
and $d_{x^{2}-y^{2}}$-wave phases (we did not found the data for
$T_{c}$ in $d_{xy}$-wave phase). We would like to emphasize that
each point in the right panel of Fig.\ref{fig:Graphene-unit-cell}
is a result of a graph similar to the left panel of Fig. \ref{fig:Hubbard}
with many points, each being a solution of self-consistency equations
\eqref{eq:self-cons}. This proves the solidity of our benchmark.

\begin{figure}
\includegraphics[viewport=24bp 0bp 422bp 686bp,clip,angle=270,width=0.5\columnwidth]{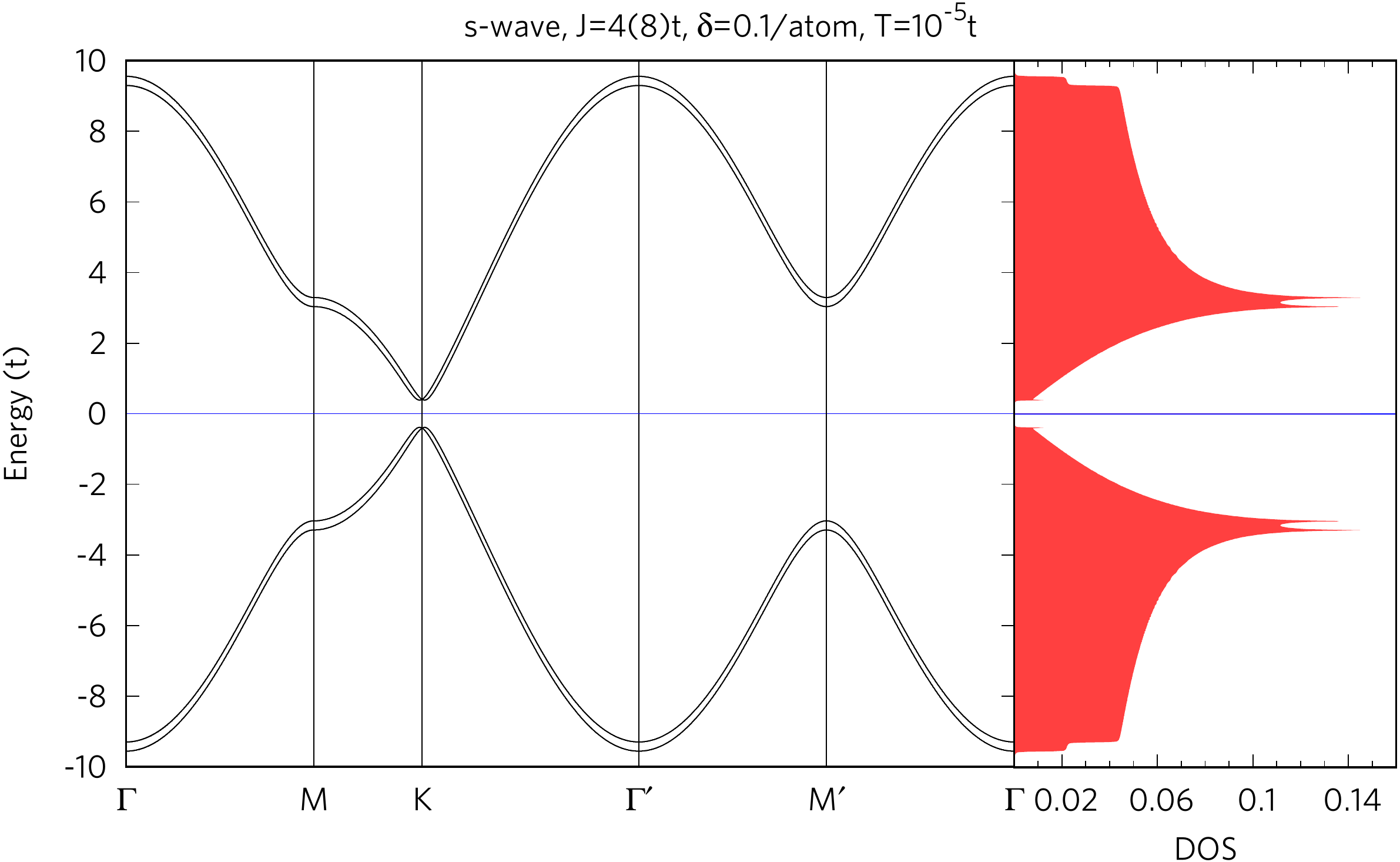}\includegraphics[viewport=24bp 0bp 423bp 693bp,clip,angle=270,width=0.5\columnwidth]{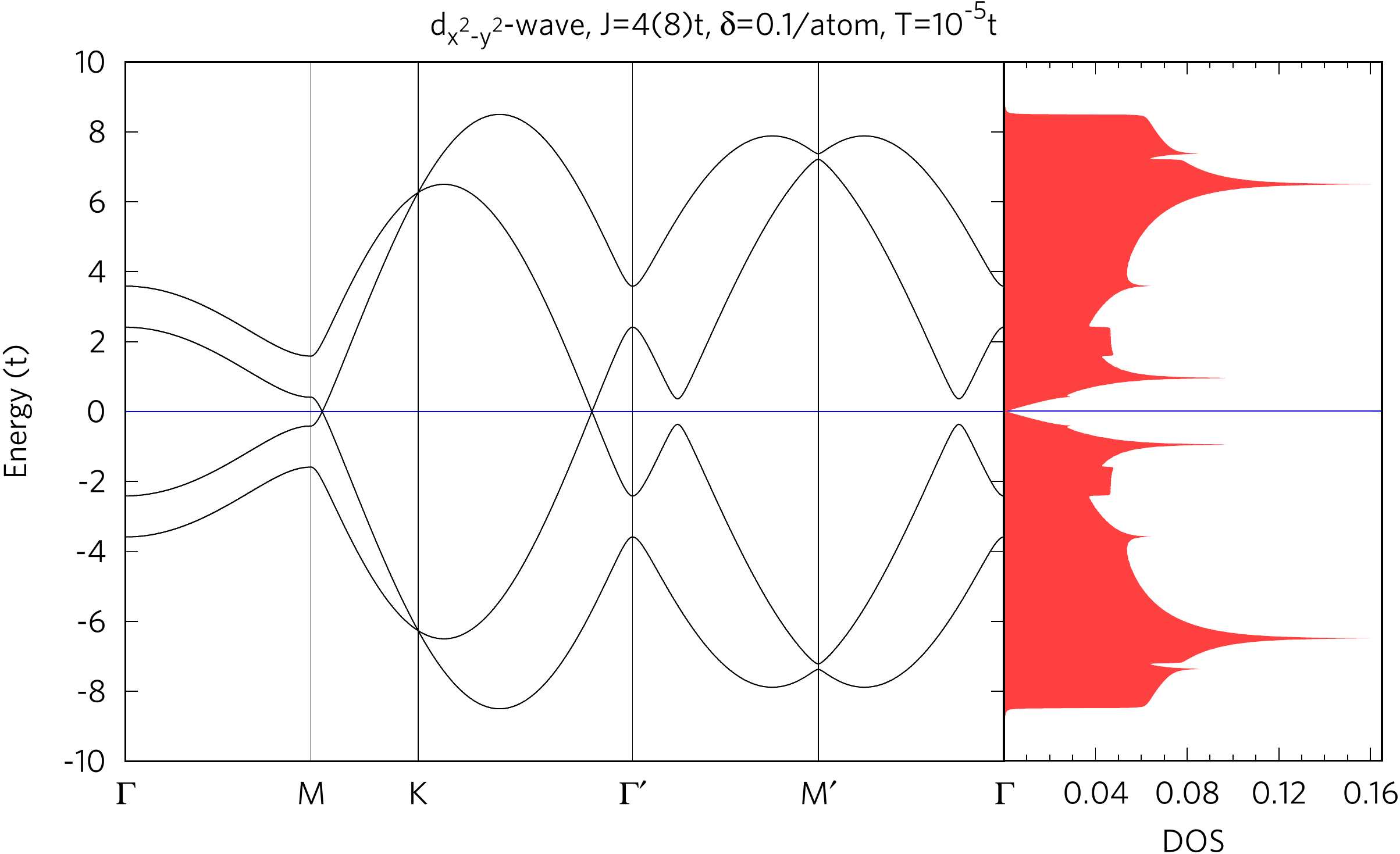}

\caption{\label{fig:DOS_Gra}Left panel: band structure along the path $\Gamma\to M\to K\to\Gamma^{\prime}\to M^{\prime}\to\Gamma$
and DOS for the $s$-wave phase of graphene. $\delta=0.1$/atom, $T=10^{-5}t$
. $\Gamma^{\prime}$ point is the center of the adjacent BZ with the
coordinates $(1,0).$ Right panel: Band structure along the path $\Gamma\to M\to K\to\Gamma^{\prime}\to M^{\prime}\to\Gamma$
and DOS for $d_{x^{2}-y^{2}}$-wave phase of graphene. $\eta=0.1$/atom,
$T=10^{-5}t$ . $\Gamma^{\prime}$ point is the center of the adjacent
BZ with the coordinates $(1,0).$ Notice that in this case, a difference
appears between $K$ and $K^{\prime}$ as well as between $M=\pi(\frac{1}{2},\frac{1}{2})$
and $M^{\prime}=\pi(\frac{1}{2},0)$.}
\end{figure}
\begin{figure}
\includegraphics[angle=270,width=0.5\columnwidth]{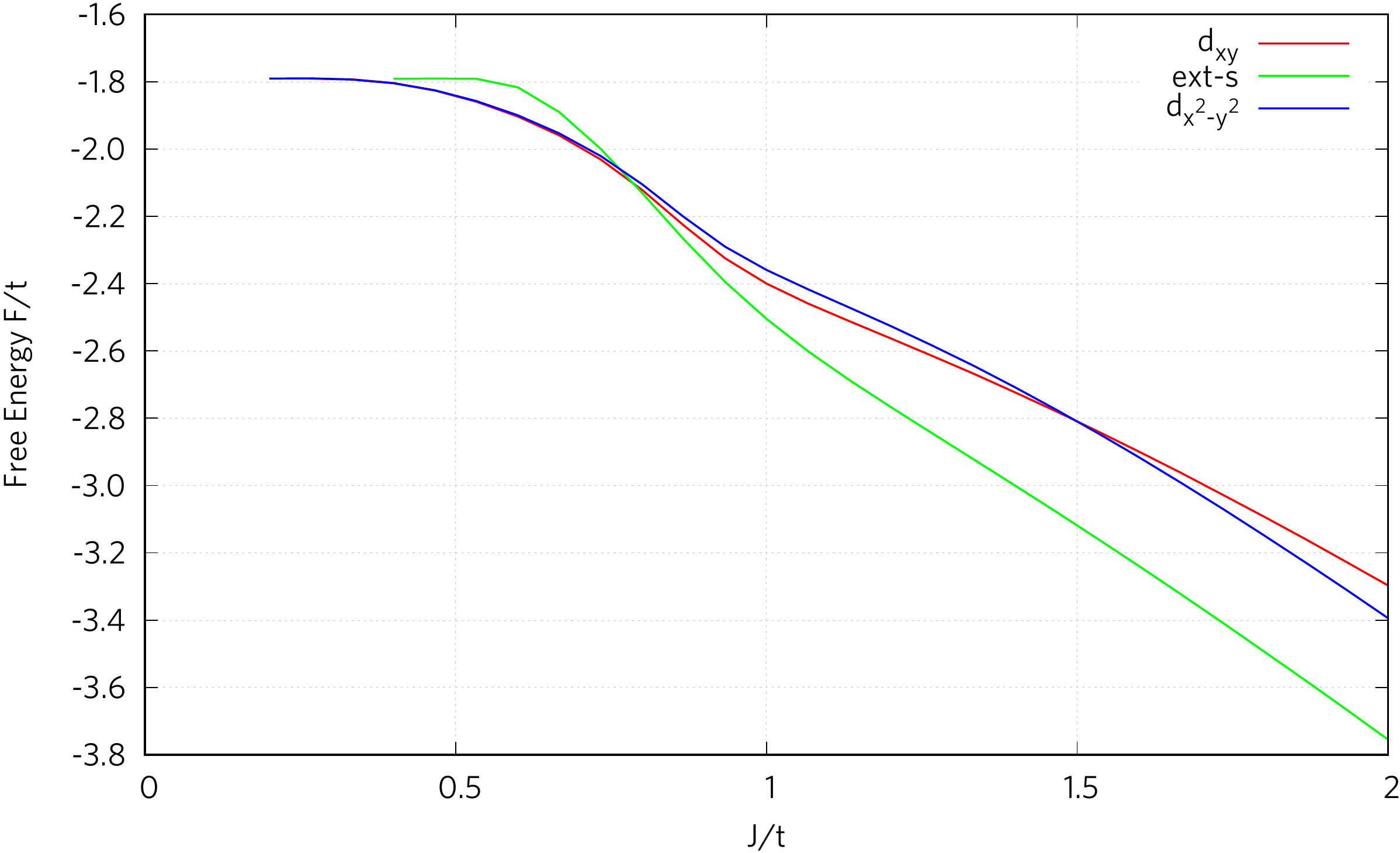}

\caption{\label{fig:E_vs_J_gra}Comparison of the $J$-dependence of the free
energy for $s$-, $d_{x^{2}-y^{2}}$- and $d_{xy}$-wave phases of
graphene with the $J$ definition compatible with the Ref.\citep{Black2014}.
Doping is fixed at $\eta=0.1$ per C atom. Superconducting solutions
do not to exist for small $J$; the respective SC order parameters
flow to zero, so that the system transits to a Fermi-liquid solution
with the free energy independent on $J$ ($F(J)\approx-1.8t$).}
\end{figure}

To have an insight into the electronic structure of these two solutions,
we compare the density of states (DOS) and the band structure in the
extended-$s$ and $d_{x^{2}-y^{2}}$-wave phases, as shown in Fig.\ref{fig:DOS_Gra}.
One can see that in the former case, there is a full gap in the DOS,
while in the latter case DOS goes to zero linearly in the vicinity
of $\omega=0$ as expected for $d_{x^{2}-y^{2}}$-symmetry gap. In
the $k$-space, the true gap opens at $K$ (and $K^{\prime}$) points
for the $s$-wave phase, while for the $d_{x^{2}-y^{2}}$-wave case
there are only two nodal points on the line $M\to K\to\Gamma^{\prime}$
whose exact position depends on doping $\eta$ and $J$. Notice that
in the latter case, there appears a difference between $K$ and $K^{\prime}$
as well as between $M=\pi(\frac{1}{2},\frac{1}{2})$ and $M^{\prime}=\pi(\frac{1}{2},0)$.
Peculiarly, in the latter case the linear spectrum, intrinsic to $H_{0}$
is restored in the $d_{x^{2}-y^{2}}$-wave solution, although bearing
a different meaning of Cooper pair nodal quasi-particles. In addition,
the overall appearance of the band structure is very different in
the two cases: in the extended-$s$ case $k$-dependence of the bands
is that of the non-interacting dispersion (although renormalized by
a coefficient proportional to $J$), which is also the gap function
in this case, while in the $d_{x^{2}-y^{2}}$-wave case the overall
$k$-dependence is a result of both non-interacting dispersion and
the corresponding gap function. Such a big difference in the band
structure leads to a sizable change in the total free energy among
the phases. In Fig.\ref{fig:E_vs_J_gra}, we show the comparison of
the free energy as a function of $J$ the three superconducting phases.
The extended-$s$ phase is the lowest one at $J>0.75t$ (in the notations
of Ref.\citep{Black2007}), while for $J<0.75t$ the $d$-wave phases
(almost indistinguishable between each other in free energy) are most
favorable. 

We emphasize that the symmetry classification, derived in Ref.\citep{Black2007},
is only rigorous in the proximity of the transition temperature, since
the linearized self-consistency equations were used. Another validity
condition for this classification is neglecting the band renormalization
as explained above. Treatment of the full interaction term would introduce
additional non-superconducting order parameters and would complicate
significantly the symmetry analysis. Under the conditions of Ref.
\citep{Black2007}, we have found that for the extended-$s$ and $d_{x^{2}-y^{2}}$
phases the respective symmetry relations hold exactly in the whole
range of their existence, while for the $d_{xy}$ phase the symmetry
relations are only asymptotic in the limit $T\to T_{c}$. By this
we show the workability of the $\Theta\Phi$ for a multi-band model
and accessibility of the BCS states of different symmetry in it together
with their temperature dependence.

\subsection{Magnetic phases of iron}

The magnetism of iron is of vital fundamental and technological importance
and was extensively studied (see \emph{e.g.} Ref. \citep{pepperhoff2001}
for review and references therein as well as Ref.\citep{Marsman_2002,Hobbs_2003}).
Among various iron phases we chose $\alpha$- and $\gamma$-Fe in
order to benchmark $\Theta\Phi$. In $\alpha$-Fe we compared the
FM phase as obtained in $\Theta\Phi$ and DFT+\emph{U} calculations
in VASP, while in $\gamma$-Fe after comparing the NM and FM phase
with VASP DFT+\emph{U}, with use of $\Theta\Phi$ we addressed the
AFM phase as well and showed that the system in this phase gains the
energy with respect to the FM one.

We have performed DFT simulations using \textsc{vasp} \citep{vasp}
in the PBE/GGA \citep{gga} version. The energy cutoff was set at
500 eV. In the DFT+\emph{U} calculations, $U=9$ eV, $J=1$ eV within
Dudarev's formulation were used\citep{Dudarev98}. 

The effective tight-binding Hamiltonian was obtained by \textsc{wannier90}
\citep{wannier90}, with the frozen window method for disentanglement
with inner window of $[-2.61:21]$ eV and the outer one of $[-2.61:70]$
eV. Local orbitals of $s,p$ and $d$ character were included and
the final Hamiltonian, expressed in the maximally localized Wannier
orbitals basis was used as input for $\Theta\Phi$. The Hubbard-$U$
correction terms where applied to Fe $d$ orbitals both in DFT and
$\Theta\Phi$ settings. We have used the same values of $U$ and $J$
for comparison with \textsc{vasp}. In order to tackle the problem
of double counting, typical in post-DFT approaches, we implemented
in $\Theta\Phi$ the modified Coulomb term of the form \citep{Czyzyk_1994}:
\begin{align}
H_{U} & =\frac{1}{2}\sum_{l,l^{\prime},r,\sigma}U_{ll^{\prime}}\left(n_{l,r,\sigma}-\left\langle n_{l,r,\sigma}^{0}\right\rangle \right)\left(n_{l^{\prime},r,-\sigma}-\left\langle n_{l^{\prime},r,-\sigma}^{0}\right\rangle \right),\label{eq:altHU}\\
 & +\frac{1}{2}\sum_{l,l^{\prime},r,\sigma}\left(U_{ll^{\prime}}-J_{ll^{\prime}}\right)\left(n_{l,r,\sigma}-\left\langle n_{l,r,\sigma}^{0}\right\rangle \right)\left(n_{l^{\prime},r,\sigma}-\left\langle n_{l^{\prime},r,\sigma}^{0}\right\rangle \right),\nonumber 
\end{align}
\begin{figure*}
\begin{minipage}[c]{0.5\textwidth}%
\subfloat[\foreignlanguage{english}{~$\Theta\Phi$}]{\includegraphics[viewport=20bp 60bp 550bp 550bp,clip,width=1\textwidth]{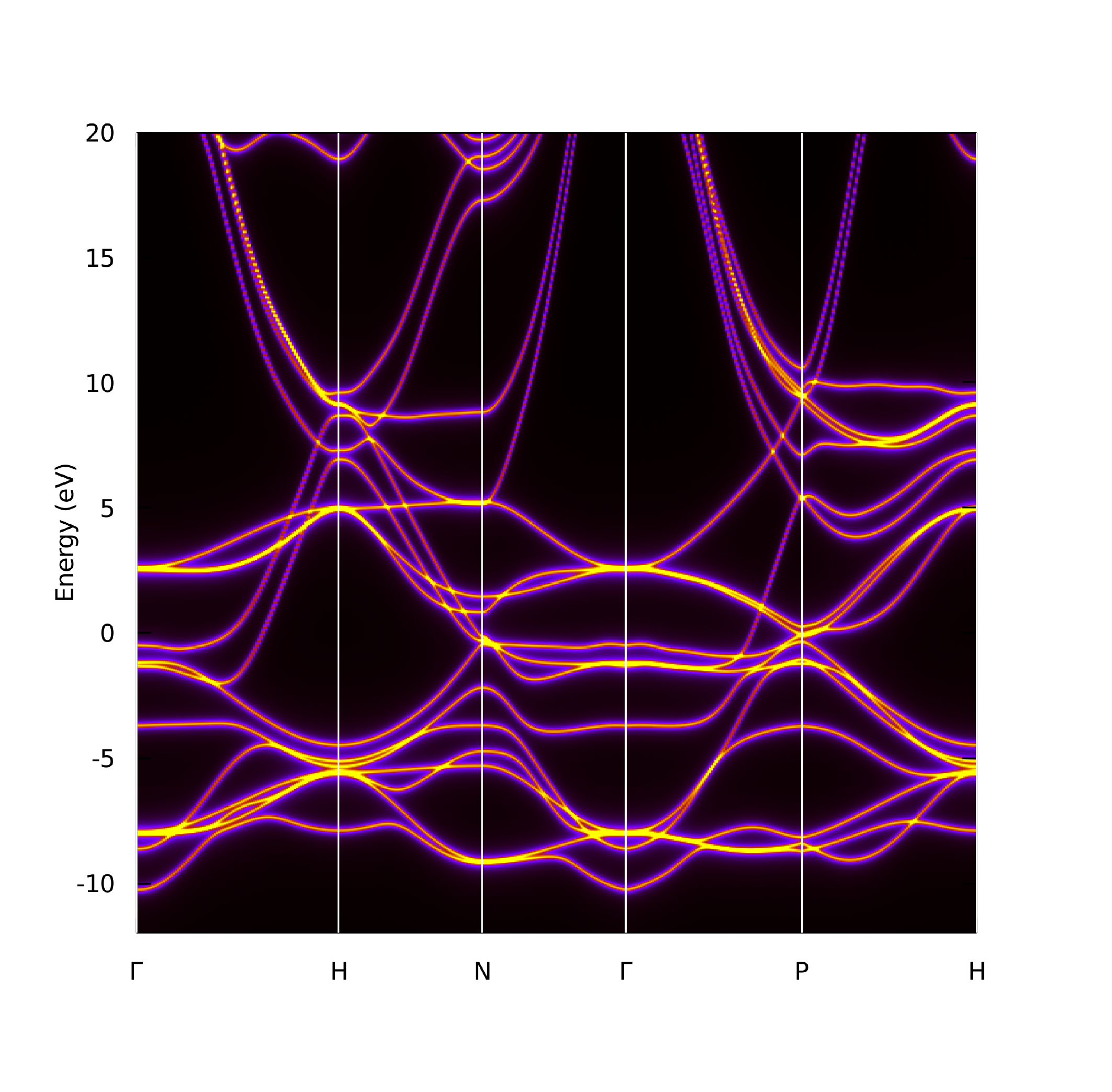}

}%
\end{minipage}\hfill{}%
\begin{minipage}[c]{0.5\textwidth}%
\subfloat[VASP]{\includegraphics[viewport=0bp 0bp 676bp 610bp,clip,width=1\textwidth]{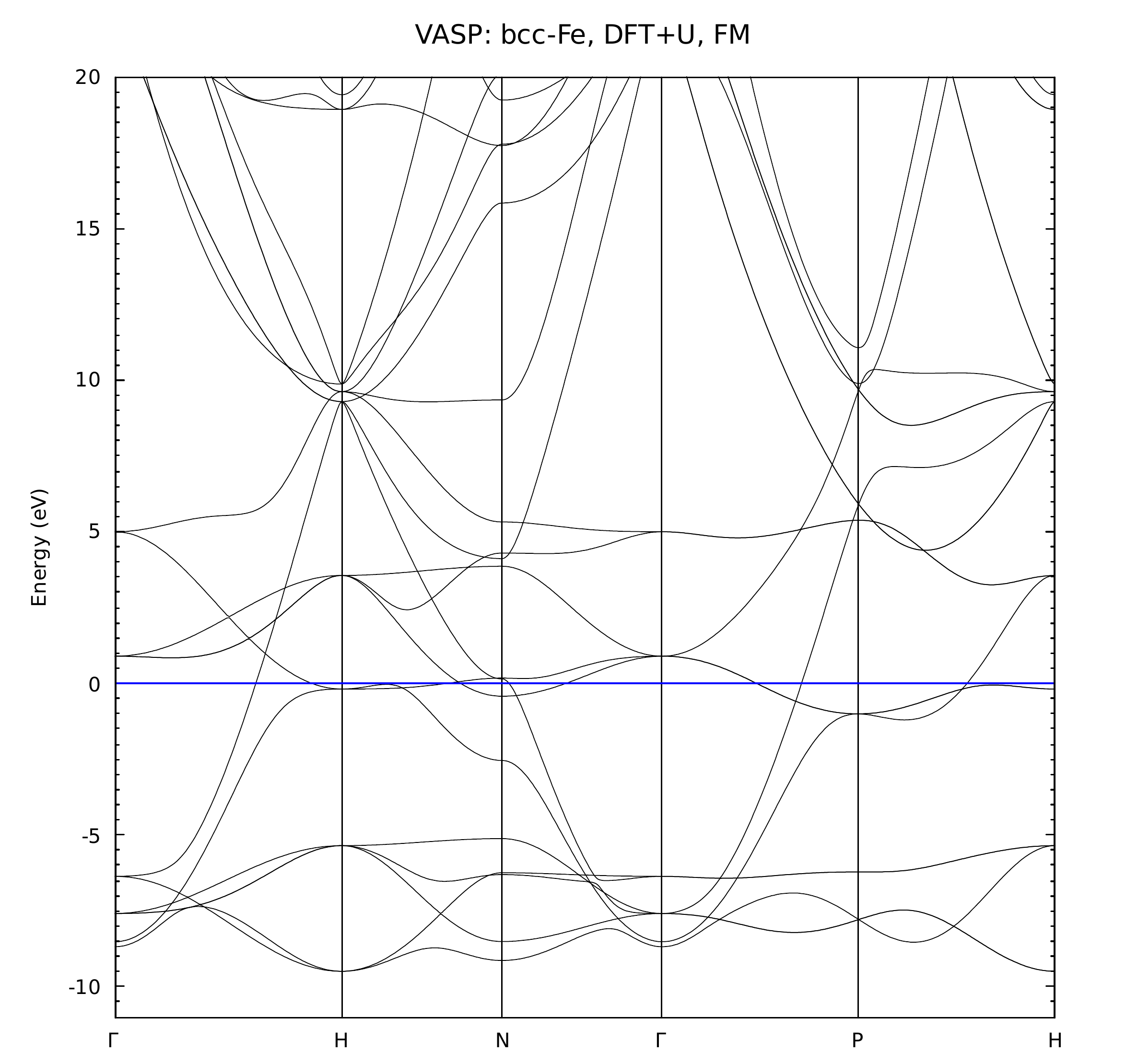}

}%
\end{minipage}

\caption{\label{fig:BandsaFe}Band structure comparison between $\Theta\Phi$
(left panel) and VASP (right panel) for ferromagnetic calculations
in $\alpha$-Fe.}
\end{figure*}
\begin{table}
\begin{tabular}{|c|c|c|c|}
\hline 
 & NM & FM ($\Theta\Phi$)  & FM(VASP)\tabularnewline
\hline 
\hline 
$\Delta E_{tot}$ (eV) & $0$ & $-6.4$ & $-5.0$\tabularnewline
\hline 
$\mu$ ($\mu_{B}$) & $0$ & $\phantom{-}3.1$ & $\phantom{-}3.2$\tabularnewline
\hline 
\end{tabular}\caption{\label{tab:TabaFe}Total energy gain $E_{tot}$ and Fe magnetic moment
comparison between $\Theta\Phi$ and VASP for ferromagnetic calculations
in $\alpha$-Fe.}
\end{table}
which has both $U_{ll^{\prime}}$ and $J_{ll^{\prime}}$ terms orbital
dependent while $\left\langle n_{l,r,\sigma}^{0}\right\rangle $ are
average occupations of the ``correlated'' orbitals as coming from
DFT. Rewritten this way, the Hubbard $U$ term describes the electronic
fluctuations around the DFT values, without shifting the center of
mass of a ``correlated'' band as a whole. From the practical side,
the interaction term eq.\eqref{eq:altHU} can be easily ``assembled''
from the elemental bits defined in \ref{sec:Appendix-A}. For the
sake of comparison with VASP we used isotropic $U_{jj^{\prime}}=U$
and $J_{jj^{\prime}}=J$, although $\Theta\Phi$ can handle arbitrary
interaction matrices.

A few words should be said about which terms contribute to the Wick's
theorem decoupling of $H_{U}$ in $\Theta\Phi$ as compared to DFT+\emph{U}.
In the former, we retain all the terms in eq.\eqref{eq:HUMF}, while
in DFT+\emph{U} only the first, ``decoupled'' term is retained \citep{Anisimov_1991,Anisimov_1997}.
Even if the ground state without ``anomalous'' averages is considered,
there are always sizable terms, contributing to the inter-orbital
hopping, missing in traditional DFT+\emph{U}. Our treatment, therefore,
is more complete and consistent.

For $\alpha$-Fe we used the bcc lattice with $a=2.858$\AA\; and
a $8\times8\times8$ Monkhorst-Pack $k$-point mesh \citep{mp} (both
in $\Theta\Phi$ and VASP). In this case we considered a ferromagnetic
solution (FM) and a non-magnetic solution (NM) and used the energy
of the latter as a reference point for comparing the relative stability
of the former. It can be seen from Table \ref{tab:TabaFe} that the
agreement between $\Theta\Phi$ and VASP is in general good: FM state
gains the energy with respect to the NM one and the $d$-shell magnetic
moment is very similar. The energy gain in $\Theta\Phi$ is approximately
$30\%$ greater, due to correct treatment of the intra-atomic orbital
hopping terms as explained above. The comparison of the band structures
is presented in Figure\ref{fig:BandsaFe}. 

For $\gamma$-Fe we considered the fcc lattice with $a=3.583$\AA\;
and a $8\times8\times8$ Monkhorst-Pack $k$-point mesh \citep{mp}.
In addition to the FM and NM states defined above, in this case we
also consider an antiferromagnetic state (AFM) with the pitch vector
$\mathbf{Q}=\pi(0,1,1)$ in terms of the vectors of the reciprocal
lattice given relative to the primitive cell, which corresponds to
an alternation of oppositely magnetized ferromagnetic planes along
the $z$-axis of the conventional cubic unit cell.

\begin{figure*}
\begin{minipage}[c]{0.54\textwidth}%
\subfloat[~$\Theta\Phi$]{\includegraphics[viewport=30bp 55bp 610bp 510bp,clip,width=1\textwidth]{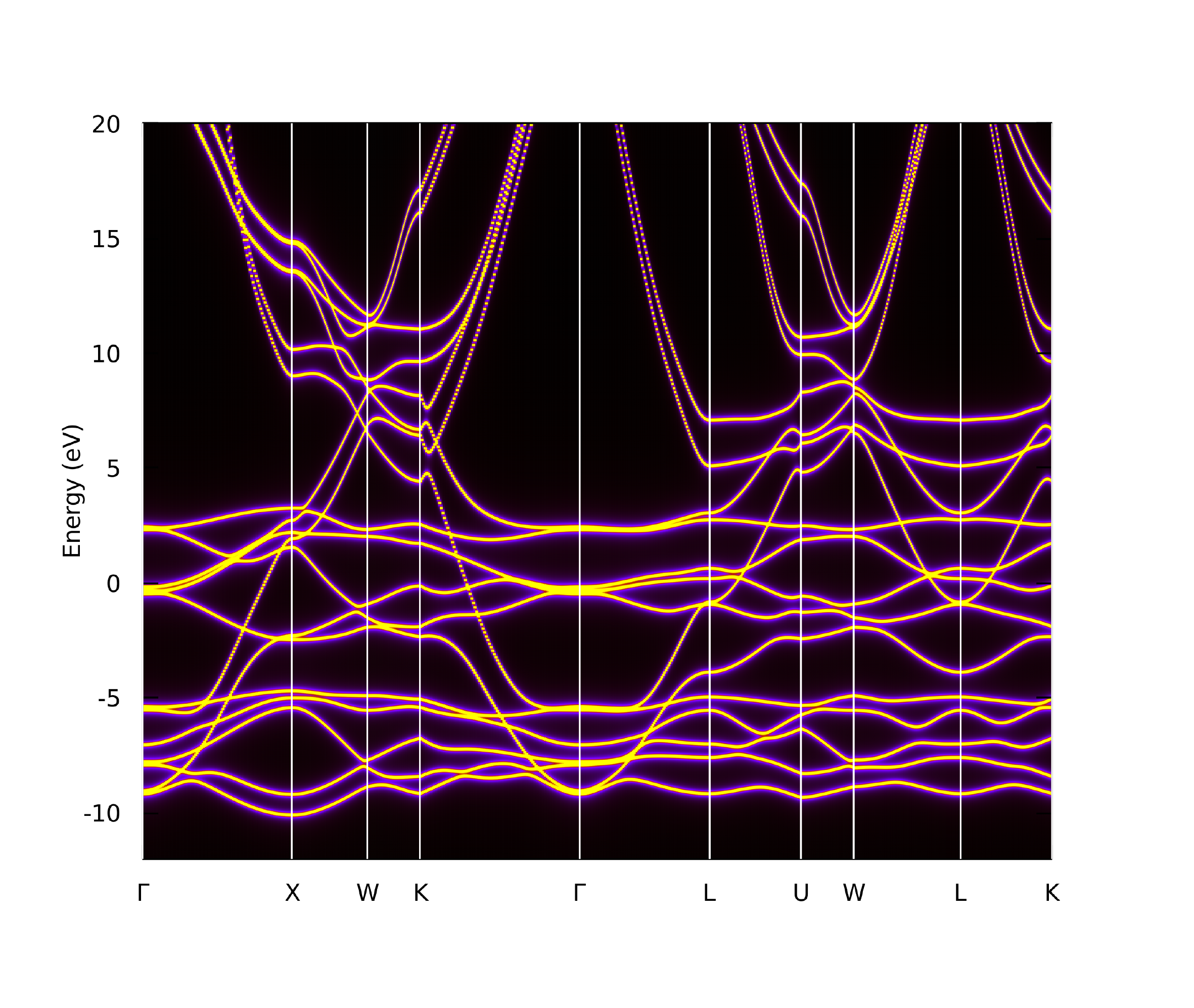}

}%
\end{minipage}\hfill{}%
\begin{minipage}[c]{0.44\textwidth}%
\subfloat[\foreignlanguage{american}{VASP}]{\includegraphics[viewport=60bp 0bp 740bp 640bp,clip,width=1\textwidth]{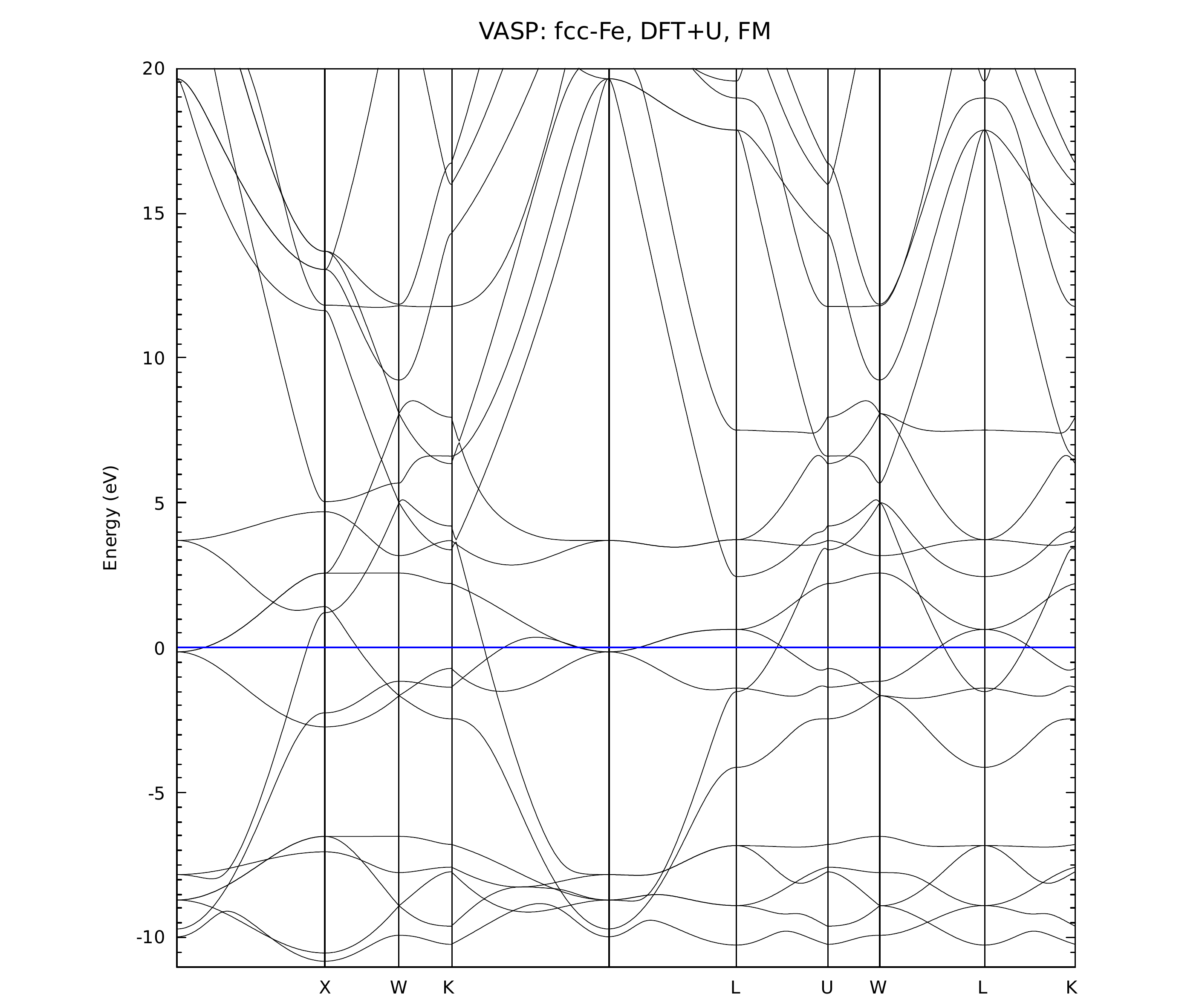}

}%
\end{minipage}

\caption{\foreignlanguage{american}{\label{fig:BandsgFe}Band structure comparison between $\Theta\Phi$
(left panel) and VASP (right panel) for ferromagnetic calculations
in $\gamma$-Fe.}}
\end{figure*}

We found that in $\gamma$-Fe the NM state has the highest energy
among all the states considered. It can be seen from Table \ref{tab:tabgFe},
that among the magnetic states, the AFM spiral state with the above
value of $\mathbf{Q}$ is lower than the FM one by $1.6$ eV, which
is larger than the corresponding VASP figure ($0.6$ eV). The FM state
is $1.8$ eV higher than the NM one in $\Theta\Phi$, as compared
to VASP, while for the AFM state this offset amounts to $0.8$ eV.
This follows that the AFM state description in $\Theta\Phi$ is somehow
closer to VASP as compared to the description of the FM state. Such
a difference in energetics between $\Theta\Phi$ and VASP can be ascribed
to the different treatment of the Hubbard term mean-field decoupling
as explained above. Finally, the $d$-orbital magnetic moment in $\Theta\Phi$
is comparable to the VASP one. The very good comparison of the band
structure results for FM $\gamma$-Fe is presented in Figure \ref{fig:BandsgFe}.

\selectlanguage{english}%
\begin{table}
\begin{tabular}{|c|c|c|c|c|c|}
\hline 
\selectlanguage{american}%
\selectlanguage{english}%
 & \selectlanguage{american}%
NM\selectlanguage{english}%
 & \selectlanguage{american}%
FM ($\Theta\Phi$) \selectlanguage{english}%
 & \selectlanguage{american}%
AFM ($\Theta\Phi$) \selectlanguage{english}%
 & \selectlanguage{american}%
FM(VASP)\selectlanguage{english}%
 & \selectlanguage{american}%
AFM (VASP)\selectlanguage{english}%
\tabularnewline
\hline 
\hline 
\selectlanguage{american}%
$\Delta E_{tot}$ (eV)\selectlanguage{english}%
 & \selectlanguage{american}%
$0$\selectlanguage{english}%
 & \selectlanguage{american}%
$-2.9$\selectlanguage{english}%
 & \selectlanguage{american}%
$-4.5$\selectlanguage{english}%
 & \selectlanguage{american}%
$-4.7$\selectlanguage{english}%
 & \selectlanguage{american}%
$-5.3$\selectlanguage{english}%
\tabularnewline
\hline 
\selectlanguage{american}%
$\mu$ (d-shell, $\mu_{B}$)\selectlanguage{english}%
 & \selectlanguage{american}%
$0$\selectlanguage{english}%
 & \selectlanguage{american}%
$\phantom{-}3.2$\selectlanguage{english}%
 & \selectlanguage{american}%
$\phantom{-}3.3$\selectlanguage{english}%
 & \selectlanguage{american}%
$\phantom{-}3.6$\selectlanguage{english}%
 & \selectlanguage{american}%
$\phantom{-}3.5$\selectlanguage{english}%
\tabularnewline
\hline 
\end{tabular}\caption{\foreignlanguage{american}{\label{tab:tabgFe}Total energy gain $\Delta E_{tot}$ and Fe $d$-shell
magnetic moment comparison between $\Theta\Phi$ and VASP for ferromagnetic
and anti-ferromagnetic $\left[\mathbf{Q}=\pi(0,1,1)\right]$ calculations
in $\gamma$-Fe.}}
\end{table}

\selectlanguage{american}%

\section{CONCLUSIONS, AND PERSPECTIVES \label{sec:Discussion-and-Conclusions}}

As we mentioned in the Introduction, the available software possesses
a limited functionality with respect to the possible magnetic superstructures.
The generally implemented technology of multiplying the chemical unit
cells allows to mimic only commensurate magnetic structures and even
the simplest of those, since increasing the unit cell size by a factor
of $n$ leads to the $n^{3}$ increase of the required computational
resources due to the requirements of the diagonalization procedures. 

On examples of two model Hamiltonians - the Hubbard one for one-dimensional
chain and the $t$-$J$ one for graphene we demonstrate the capacity
of the package to reproduce the temperature dependent BCS solutions
in these models. The characteristic features of the obtained dependency
of the BCS order parameters known from analytical theory are reproduced.
Since the tight-binding Hamiltonians are the inputs for the proposed
procedure, the ports to the sources of such are developed. With the
use of the tight-binding parameters as extracted from the VASP calculation
projected on the $3d4s4p$ local basis the band structures of the
ferromagnetic bcc and antiferromagnetic fcc iron as well as the respective
magnetic momenta and the relative energies with respect to nonmagnetic
phases are fairly reproduced. The antiferromagnetic ordered state
is described by imposing the experimentally known value of the superstructure
wave vector, without extending the chemical unit cell. By this the
applicability of the code to the multi-band models of interacting
electrons in solids is demonstrated. Moreover, there exists a close
relation\citep{RVB_review_2017} between the BCS state and the resonating
valence bonds state, introduced in the solid state context \citep{Fazekas_phil}
as an option for frustrated antiferromagnets and later hypothesized
to be the state of high-T$_{c}$ cuprate superconductors \citep{Anderson_1987}.
Owing to this relation, our package $\Theta\varPhi$ is ready for
direct simulation of such states too.

\section*{ACKNOWLEDGMENTS}

This work is partially supported by DFG ``Stahl - ab initio'' Sonderforschungsbereich
761 and by the Volkswagenstiftung (grant \textnumero{} 151110 ``Deductive
Quantum Molecular Mechanics of Carbon Allotropes'' in the frame of
the Initiative of Trilateral Partnership Cooperation Projects between
Scholars and Scientists from The Ukraine, Russia, and Germany).

\appendix

\section{\label{sec:Appendix-A}Mean-field decoupling}

As mentioned above, the \emph{decoupling} serves to produce the Fockian
(mean-field Hamiltonian) from the original Hamiltonian and a density
matrix. In the original Hamiltonian the electron-electron interactions
(e.g. the Coulomb ones) enter through the products of four Fermi operators
$c_{1}^{\dagger}c_{2}c_{3}^{\dagger}c_{4}$. The decoupling of the
products of four operators is based on Wick's theorem\citep{Wick_1950}
and replaces them by an expression containing products of only two
operators complemented by the \emph{averages} $\left\langle ...\right\rangle $
of the product of two remaining operators - the elements of the density
matrix used for decoupling: 
\begin{align}
c_{1}^{\dagger}c_{2}c_{3}^{\dagger}c_{4} & \Rightarrow\left\langle c_{1}^{\dagger}c_{2}\right\rangle c_{3}^{\dagger}c_{4}+c_{1}^{\dagger}c_{2}\left\langle c_{3}^{\dagger}c_{4}\right\rangle -\left\langle c_{1}^{\dagger}c_{2}\right\rangle \left\langle c_{3}^{\dagger}c_{4}\right\rangle \nonumber \\
 & -\left\langle c_{1}^{\dagger}c_{4}\right\rangle c_{3}^{\dagger}c_{2}-c_{1}^{\dagger}c_{4}\left\langle c_{3}^{\dagger}c_{2}\right\rangle +\left\langle c_{1}^{\dagger}c_{4}\right\rangle \left\langle c_{3}^{\dagger}c_{2}\right\rangle \label{eq:InteractionToEOMMatrix}\\
 & +\left\langle c_{1}^{\dagger}c_{3}^{\dagger}\right\rangle c_{4}c_{2}+c_{1}^{\dagger}c_{3}^{\dagger}\left\langle c_{4}c_{2}\right\rangle -\left\langle c_{1}^{\dagger}c_{3}^{\dagger}\right\rangle \left\langle c_{4}c_{2}\right\rangle \nonumber 
\end{align}
(the products of the two averages - the last terms in each row are
serving to make the mean value of the mean-field Hamiltonian equal
to the mean value of the original Hamiltonian). In the standard software
only first two rows of eq.\eqref{eq:InteractionToEOMMatrix} are implemented,
setting the \emph{anomalous} averages $\left\langle c_{1}^{\dagger}c_{3}^{\dagger}\right\rangle $
and $\left\langle c_{4}c_{2}\right\rangle $ identically equal to
zero. These averages are characteristic for the BCS and RVB solutions
of the SCF electronic problem and are not generally vanishing in our
setting. The presence of the non-vanishing anomalous averages breaks
an important symmetry tacitly assumed in all quantum chemistry software:
the conservation of the number of particles. As a consequence, non-vanishing
matrix elements between the electron and hole states appear in the
mean-field Hamiltonian.

Here we report the mean-field reduction of the Hamiltonian terms $H_{U}$,
$H_{V}$ and $H_{J}$. We begin with $H_{U}$. The most general form
of $H_{U}$ considered here reads as follows:
\begin{equation}
H_{U}(s,s^{\prime})=\sum_{l,l^{\prime},r}U_{l,l^{\prime}}^{s,s^{\prime}}c_{l,r,s}^{\dagger}c_{l,r,s}^{\phantom{\dagger}}c_{l^{\prime},r,s^{\prime}}^{\dagger}c_{l^{\prime},r,s^{\prime}}^{\phantom{\dagger}}=\sum_{l,l^{\prime},r}U_{l,l^{\prime}}^{s,s^{\prime}}n_{l,r,s}n_{l^{\prime},r,s^{\prime}}.
\end{equation}
 The average value is:

\begin{align}
\left\langle H_{U}(s,s^{\prime})\right\rangle  & =\sum_{l,l^{\prime},r}U_{l,l^{\prime}}^{s,s^{\prime}}\left\{ \left\langle n_{l,r,s}\right\rangle \left\langle n_{l^{\prime},r,s^{\prime}}\right\rangle \right.\nonumber \\
 & -\left\langle c_{l,r,s}^{\dagger}c_{l^{\prime},r,s^{\prime}}^{\phantom{\dagger}}\right\rangle \left\langle c_{l^{\prime},r,s^{\prime}}^{\dagger}c_{l,r,s}^{\phantom{\dagger}}\right\rangle \delta_{s^{\prime},s}\\
 & \left.+\left\langle c_{l,r,s}^{\dagger}c_{l^{\prime},r,s^{\prime}}^{\dagger}\right\rangle \left\langle c_{l^{\prime},r,s^{\prime}}^{\phantom{\dagger}}c_{l,r,s}^{\phantom{\dagger}}\right\rangle \delta_{s^{\prime},-s}\right\} .\nonumber 
\end{align}
Here the first term is the ``uncoupled'' one, the second term describes
the inter-orbital hopping (Coulomb exchange), while the last one is
due to superconducting fluctuations.

The ``linearized'' mean-field term becomes:

\begin{align}
H_{U}^{MF}(s,s^{\prime}) & =\sum_{l,l^{\prime},r}U_{l,l^{\prime}}^{s,s^{\prime}}\left\{ \left\langle n_{l,r,s}\right\rangle n_{l^{\prime},r,s^{\prime}}+\left\langle n_{l^{\prime},r,s^{\prime}}\right\rangle n_{l,r,s}\right.\label{eq:HUMF}\\
 & -\left(\left\langle c_{l,r,s}^{\dagger}c_{l^{\prime},r,s^{\prime}}^{\phantom{\dagger}}\right\rangle c_{l^{\prime},r,s^{\prime}}^{\dagger}c_{l,r,s}^{\phantom{\dagger}}+\left\langle c_{l^{\prime},r,s^{\prime}}^{\dagger}c_{l,r,s}^{\phantom{\dagger}}\right\rangle c_{l,r,s}^{\dagger}c_{l^{\prime},r,s^{\prime}}^{\phantom{\dagger}}\right)\delta_{s^{\prime},s}\nonumber \\
 & \left.+\left(\left\langle c_{l,r,s}^{\dagger}c_{l^{\prime},r,s^{\prime}}^{\dagger}\right\rangle c_{l^{\prime},r,s^{\prime}}^{\phantom{\dagger}}c_{l,r,s}^{\phantom{\dagger}}+\left\langle c_{l^{\prime},r,s^{\prime}}^{\phantom{\dagger}}c_{l,r,s}^{\phantom{\dagger}}\right\rangle c_{l,r,s}^{\dagger}c_{l^{\prime},r,s^{\prime}}^{\dagger}\right)\delta_{s^{\prime},-s}\right\} \nonumber \\
 & =\sum_{l,l^{\prime},r}U_{l,l^{\prime}}^{s,s^{\prime}}\left\{ \rho_{lr,lr}(0)n_{l,r,s}+\rho_{l^{\prime}s^{\prime},l^{\prime}s^{\prime}}(0)n_{l,r,s}\right.\nonumber \\
 & -\left(\rho_{ls,l^{\prime}s^{\prime}}(0)c_{l^{\prime},r,s^{\prime}}^{\dagger}c_{l,r,s}^{\phantom{\dagger}}+\rho_{l^{\prime}s^{\prime},ls}(0)c_{l,r,s}^{\dagger}c_{l^{\prime},r,s^{\prime}}^{\phantom{\dagger}}\right)\delta_{s^{\prime},s}\nonumber \\
 & \left.+\left(\rho_{ls,l^{\prime}+2Ls^{\prime}}(0)c_{l^{\prime},r,s^{\prime}}^{\phantom{\dagger}}c_{l,r,s}^{\phantom{\dagger}}+\rho_{l^{\prime}+2Ls^{\prime},ls}(0)c_{l,r,s}^{\dagger}c_{l^{\prime},r,s^{\prime}}^{\dagger}\right)\delta_{s^{\prime},-s}\right\} .\nonumber 
\end{align}
Here we recall that the density matrix is a $4L\times4L$ matrix,
so that the index $\left\{ ls,l+2Ls\right\} $ signifies the particle-particle
(superconducting pairing) channel. It can be easily seen that $\left\langle H_{U}^{MF}\right\rangle =\left\langle H_{U}\right\rangle $.

We now turn to $H_{V}$. 
\begin{align*}
\left\langle H_{V}\right\rangle  & =\sum_{l,l^{\prime},r,\tau}V_{l,l^{\prime}}(\tau)\left\{ \left\langle n_{l,r}\right\rangle \left\langle n_{l^{\prime},r+\tau}\right\rangle \right.\\
 & -\sum_{s}\left\langle c_{l,r,s}^{\dagger}c_{l^{\prime},r+\tau,s}^{\phantom{\dagger}}\right\rangle \left\langle c_{l^{\prime},r+\tau,s}^{\dagger}c_{l,r,s}^{\phantom{\dagger}}\right\rangle \\
 & +\sum_{s}\left\langle c_{l,r,s}^{\dagger}c_{l^{\prime},r+\tau,-s}^{\dagger}\right\rangle \left\langle c_{l^{\prime},r+\tau,-s}^{\phantom{\dagger}}c_{l,r,s}^{\phantom{\dagger}}\right\rangle 
\end{align*}
\begin{align*}
H_{V}^{MF} & =\sum_{l,r,s}V_{l,l^{\prime}}(\tau)\left\{ \left\langle n_{l,r}\right\rangle c_{l^{\prime},r+\tau,s}^{\dagger}c_{l^{\prime},r+\tau,s}^{\phantom{\dagger}}+\left\langle n_{l^{\prime},r+\tau}\right\rangle c_{l,r,s}^{\dagger}c_{l,r,s}^{\phantom{\dagger}}\right.\\
 & -\left\langle c_{l,r,s}^{\dagger}c_{l^{\prime},r+\tau,s}^{\phantom{\dagger}}\right\rangle c_{l^{\prime},r+\tau,s}^{\dagger}c_{l,r,s}^{\phantom{\dagger}}-\left\langle c_{l^{\prime},r+\tau,s}^{\dagger}c_{l,r,s}^{\phantom{\dagger}}\right\rangle c_{l,r,s}^{\dagger}c_{l^{\prime},r+\tau,s}^{\phantom{\dagger}}\\
 & +\left.\left\langle c_{l,r,s}^{\dagger}c_{l^{\prime},r+\tau,-s}^{\dagger}\right\rangle c_{l^{\prime},r+\tau,-s}^{\phantom{\dagger}}c_{l,r,s}^{\phantom{\dagger}}+\left\langle c_{l^{\prime},r+\tau,-s}^{\phantom{\dagger}}c_{l,r,s}^{\phantom{\dagger}}\right\rangle c_{l,r,s}^{\dagger}c_{l^{\prime},r+\tau,-s}^{\dagger}\right\} \\
 & =\sum_{l,l^{\prime},r,s}V_{l,l^{\prime}}(\tau)\biggl\{\sum_{s^{\prime}}\rho_{ls^{\prime},ls^{\prime}}(0)c_{l^{\prime},r+\tau,s}^{\dagger}c_{l^{\prime},r+\tau,s}^{\phantom{\dagger}}+\sum_{s^{\prime}}\rho_{l^{\prime}s^{\prime},l^{\prime}s^{\prime}}(0)c_{l,r,s}^{\dagger}c_{l,r,s}^{\phantom{\dagger}}\\
 & +\rho_{l,s,l^{\prime}+2L,-s}(\tau)c_{l^{\prime},r+\tau,-s}^{\phantom{\dagger}}c_{l,r,s}^{\phantom{\dagger}}+\rho_{l^{\prime}+2L,-s,l,s}(-\tau)c_{l,r,s}^{\dagger}c_{l^{\prime},r+\tau,-s}^{\dagger}\\
 & -\rho_{l,s,l^{\prime},s}(\tau)c_{l^{\prime},r+\tau,\sigma}^{\dagger}c_{l,r,s}^{\phantom{\dagger}}-\rho_{l^{\prime},s,l,s}(-\tau)c_{l,r,s}^{\dagger}c_{l^{\prime},r+\tau,s}^{\phantom{\dagger}}\biggr\}.
\end{align*}
As before, $\left\langle H_{V}^{MF}\right\rangle =\left\langle H_{V}\right\rangle $.
Finally, we consider $H_{J}$ term.
\begin{align}
\left\langle H_{J}\right\rangle  & =\frac{1}{4}\sum_{l,l^{\prime}r,\tau,\alpha,\beta}J_{l,l^{\prime}}^{\alpha\beta}(\tau)\left\langle S_{l,r}^{\alpha}\right\rangle \left\langle S_{l^{\prime},r+\tau}^{\beta}\right\rangle \nonumber \\
 & +\frac{1}{4}\sum_{\substack{l,l^{\prime}r,\tau,\alpha,\beta,\\
t,t^{\prime},s,s^{\prime}
}
}J_{l,l^{\prime}}^{\alpha\beta}(\tau)\sigma_{t,t^{\prime}}^{\alpha}\sigma_{s,s^{\prime}}^{\beta}\biggl\{-\left\langle c_{l,r,t}^{\dagger}c_{l^{\prime},r+\tau,t}^{\phantom{\dagger}}\right\rangle \left\langle c_{l^{\prime},r+\tau,t^{\prime}}^{\dagger}c_{l,r,t^{\prime}}^{\phantom{\dagger}}\right\rangle \delta_{s,t^{\prime}}\delta_{s^{\prime},t}\label{eq:HJ}\\
 & +\left\langle c_{l,r,t}^{\dagger}c_{l^{\prime},r+\tau,-t}^{\dagger}\right\rangle \left\langle c_{l^{\prime},r+\tau,-t^{\prime}}^{\phantom{\dagger}}c_{l,r,t^{\prime}}^{\phantom{\dagger}}\right\rangle \delta_{s,-t}\delta_{s^{\prime},-t^{\prime}}\biggr\}.\nonumber 
\end{align}
Here, the first, term also called ``decoupled'' is merely a product
of average spins at corresponding sites and can be simplified as follows:
\[
\frac{1}{4}\sum_{l,l^{\prime}r,\tau}J_{l,l^{\prime}}^{zz}(\tau)\left\langle S_{l,r}^{z}\right\rangle \left\langle S_{l^{\prime},r+\tau}^{z}\right\rangle ,
\]
as the averages of other spin components are zero. The sums on $\alpha,\beta,t,t^{\prime},s,s^{\prime}$
with Pauli matrices can be simplified by noting that half of the Pauli
matrices elements are zero. This can be more compactly rewritten \textcolor{black}{upon
introduction of three very simple auxiliary matrices $A$, $B$ and
$C$: 
\[
A=\begin{pmatrix}1 & 0 & 0\\
0 & 1 & 0\\
0 & 0 & 0
\end{pmatrix};\quad B=\begin{pmatrix}0 & 0 & 0\\
0 & 0 & 0\\
0 & 0 & 1
\end{pmatrix};\quad C=\begin{pmatrix}0 & -i & 0\\
i & 0 & 0\\
0 & 0 & 0
\end{pmatrix}.
\]
}
\begin{align*}
\left\langle H_{J}\right\rangle  & =\frac{1}{4}\sum_{\substack{l,l^{\prime}r,\\
\tau,\alpha,\beta
}
}J_{l,l^{\prime}}^{\alpha\beta}(\tau)\left\langle S_{l,r}^{\alpha}\right\rangle \left\langle S_{l^{\prime},r+\tau}^{\beta}\right\rangle \\
 & +\frac{1}{4}\sum_{\substack{l,l^{\prime}r,\\
\tau,s,s^{\prime}
}
}\left(\mathrm{Tr}\left[J_{l,l^{\prime}}(\tau)A\right]\delta_{s,-s^{\prime}}+\mathrm{Tr}\left[J_{l,l^{\prime}}(\tau)B\right]\delta_{s,s^{\prime}}+\mathrm{Tr}\left[J_{l,l^{\prime}}(\tau)C\right]\delta_{s,-s^{\prime}}(-1)^{s}\right)\times\\
 & \biggl\{-\left\langle c_{l,r,s}^{\dagger}c_{l^{\prime},r+\tau,s}^{\phantom{\dagger}}\right\rangle \left\langle c_{l^{\prime},r+\tau,s^{\prime}}^{\dagger}c_{l,rs^{\prime}}^{\phantom{\dagger}}\right\rangle +\left\langle c_{l,r,s}^{\dagger}c_{l^{\prime},r+\tau,-s}^{\dagger}\right\rangle \left\langle c_{l^{\prime},r+\tau,-s^{\prime}}^{\phantom{\dagger}}c_{l,r,s^{\prime}}^{\phantom{\dagger}}\right\rangle \biggr\}.\\
 & =\frac{1}{4}\sum_{l,l^{\prime},r,\tau}J_{l,l^{\prime}}^{zz}(\tau)\left\langle S_{l,r}^{z}\right\rangle \left\langle S_{l^{\prime},r+\tau}^{z}\right\rangle \\
 & +\frac{1}{4}\sum_{l,l^{\prime},r,\tau}\left(J_{l,l^{\prime}}^{xx}(\tau)+J_{l,l^{\prime}}^{yy}(\tau)\right)\biggl\{-\left\langle c_{l,r,\uparrow}^{\dagger}c_{l^{\prime},r+\tau,\uparrow}^{\phantom{\dagger}}\right\rangle \left\langle c_{l^{\prime},r+\tau,\downarrow}^{\dagger}c_{l,r,\downarrow}^{\phantom{\dagger}}\right\rangle +\left\langle c_{l,r,\uparrow}^{\dagger}c_{l^{\prime},r+\tau,\downarrow}^{\dagger}\right\rangle \left\langle c_{l^{\prime},r+\tau,\downarrow}^{\phantom{\dagger}}c_{j,r,\uparrow}^{\phantom{\dagger}}\right\rangle \\
 & -\left\langle c_{l,r,\downarrow}^{\dagger}c_{l^{\prime},r+\tau,\downarrow}^{\phantom{\dagger}}\right\rangle \left\langle c_{l^{\prime},r+\tau,\uparrow}^{\dagger}c_{l,r,\uparrow}^{\phantom{\dagger}}\right\rangle +\left\langle c_{l,r,\downarrow}^{\dagger}c_{l^{\prime},r+\tau,\uparrow}^{\dagger}\right\rangle \left\langle c_{l^{\prime},r+\tau,\uparrow}^{\phantom{\dagger}}c_{l,r,\downarrow}^{\phantom{\dagger}}\right\rangle \biggr\}\\
 & +\frac{i}{4}\sum_{l,l^{\prime},r,\tau}\left(J_{l,l^{\prime}}^{xx}(\tau)-J_{l,l^{\prime}}^{yy}(\tau)\right)\biggl\{-\left\langle c_{l,r,\uparrow}^{\dagger}c_{l^{\prime},r+\tau,\uparrow}^{\phantom{\dagger}}\right\rangle \left\langle c_{l^{\prime},r+\tau,\uparrow}^{\dagger}c_{l,r,\uparrow}^{\phantom{\dagger}}\right\rangle +\left\langle c_{l,r,\uparrow}^{\dagger}c_{l^{\prime},r+\tau,\downarrow}^{\dagger}\right\rangle \left\langle c_{l^{\prime},r+\tau,\uparrow}^{\phantom{\dagger}}c_{l,r,\downarrow}^{\phantom{\dagger}}\right\rangle \\
 & -\left\langle c_{l,r,\downarrow}^{\dagger}c_{l^{\prime},r+\tau,\downarrow}^{\phantom{\dagger}}\right\rangle \left\langle c_{l^{\prime},r+\tau,\downarrow}^{\dagger}c_{l,r,\downarrow}^{\phantom{\dagger}}\right\rangle +\left\langle c_{l,r,\downarrow}^{\dagger}c_{l^{\prime},r+\tau,\uparrow}^{\dagger}\right\rangle \left\langle c_{l^{\prime},r+\tau,\downarrow}^{\phantom{\dagger}}c_{l,r,\uparrow}^{\phantom{\dagger}}\right\rangle \biggr\}\\
 & +\frac{1}{4}\sum_{l,l^{\prime},r,\tau}J_{l,l^{\prime}}^{zz}(\tau)\biggl\{-\left\langle c_{l,r,\uparrow}^{\dagger}c_{l^{\prime},r+\tau,\uparrow}^{\phantom{\dagger}}\right\rangle \left\langle c_{l^{\prime},r+\tau,\downarrow}^{\dagger}c_{l,r,\downarrow}^{\phantom{\dagger}}\right\rangle +\left\langle c_{l,r,\uparrow}^{\dagger}c_{l^{\prime},r+\tau,\downarrow}^{\dagger}\right\rangle \left\langle c_{l^{\prime},r+\tau,\downarrow}^{\phantom{\dagger}}c_{l,r,\uparrow}^{\phantom{\dagger}}\right\rangle \\
 & -\left\langle c_{l,r,\downarrow}^{\dagger}c_{l^{\prime},r+\tau,\downarrow}^{\phantom{\dagger}}\right\rangle \left\langle c_{l^{\prime},r+\tau,\uparrow}^{\dagger}c_{l,r,\uparrow}^{\phantom{\dagger}}\right\rangle -\left\langle c_{l,r,\downarrow}^{\dagger}c_{l^{\prime},r+\tau,\uparrow}^{\dagger}\right\rangle \left\langle c_{l^{\prime},r+\tau,\uparrow}^{\phantom{\dagger}}c_{l,r,\downarrow}^{\phantom{\dagger}}\right\rangle \biggr\}.
\end{align*}
Here, the discrete function $(-1)^{s}$ means $+1$ for $s=\uparrow$
and $-1$ otherwise. The expression for $H_{J}^{MF}$ is obtained
from the above formula by ``linearization'' procedure, which consists
in substitution in each product like $\left\langle c_{l,r,\uparrow}^{\dagger}c_{l^{\prime},r+\tau,\downarrow}^{\dagger}\right\rangle \left\langle c_{l^{\prime},r+\tau,\downarrow}^{\phantom{\dagger}}c_{l,r,\uparrow}^{\phantom{\dagger}}\right\rangle $
with an operator of the form 
\[
\left\langle c_{l^{\prime},r+\tau,\downarrow}^{\phantom{\dagger}}c_{l,r,\uparrow}^{\phantom{\dagger}}\right\rangle c_{l,r,\uparrow}^{\dagger}c_{l^{\prime},r+\tau,\downarrow}^{\dagger}+\left\langle c_{l,r,\uparrow}^{\dagger}c_{l^{\prime},r+\tau,\downarrow}^{\dagger}\right\rangle c_{l^{\prime},r+\tau,\downarrow}^{\phantom{\dagger}}c_{l,r,\uparrow}^{\phantom{\dagger}}.
\]
The resulting expression is somewhat lengthy and is not reported here
for brevity, but fully implemented in $\Theta\Phi$. As always happens
with the mean-field ``linearization'' $\left\langle H_{J}^{MF}\right\rangle =\left\langle H_{J}\right\rangle $.

\section{\label{sec:Rotations}Spin rotations of the Hamiltonian terms}

As it is noticed in the main text, in order to describe magnetic ordering
we employ the unit-cell dependent spin quantization axes. It is not
necessary, however, to assume that the spins \emph{within} a unit
cell are all quantized along the same axis. That is to say that one
can assign to each orbital in the unit cell its own rotation matrix
determining the direction of its quantization axis in the global ``laboratory''
frame. Let $SU(2)$-matrices $\Omega(i)$ and $\Omega(j)$ be those
which rotate the spin quantization axes of the $i$-th and $j$-th
orbitals in the unit cell. Then for the pair of such orbitals when
the $j$-th orbital is located in the unit cell shifted by the lattice
vector $\mathbf{\tau}$ the matrix multiplier 
\begin{equation}
\Omega^{\dagger}(i)\Omega(\mathbf{n,\tau,Q})\Omega(j)\label{eq:GeneralSpinRotationMAtrix}
\end{equation}
must be inserted between the fermion-vector multipliers representing
the electron being destroyed in the $j$-th orbital and one being
created in the $i$-th orbital. This matrix is as well of the form
given by eq.\eqref{eq:SU2Matrix}.

Here we report without derivation the formulae for the transformation
of the matrix elements of the Hamiltonian terms involved in $\Theta\Phi$.
To begin with, we note that the extended Coulomb interaction $H_{V}$
only depends on the total occupation of a given orbital \emph{e.g.}
$n_{ir}$ which in turn is invariant with respect to the spin quantization
axis rotations, therefore so does $H_{V}$.

The on-site Hubbard one $H_{U}(\sigma,\sigma^{\prime})$ depends on
individual spin component occupation operators $n_{ir\sigma}$ and
hence, in principle, is not invariant under spin rotations, however,
only spin rotationally invariant Hubbard term (after summation on
spin indices) has physical meaning, which in turn is determined by
the form of the matrix $U_{j,j^{\prime}}^{\sigma,\sigma^{\prime}}$.
Therefore, only such form of matrices, like in Ref. \citep{Czyzyk_1994}
is used in $\Theta\Phi$. 

The most general for of the kinetic energy can be written as:
\[
\sum_{\substack{i,j,r\\
\tau,s,s^{\prime}
}
}c_{irs}^{\dagger}t_{i,j}^{ss^{\prime}}(\tau)c_{jr+\tau s^{\prime}}^{\phantom{\dagger}},
\]
where the tensor $t_{i,j}^{ss^{\prime}}(\tau)$ defines kinetic energy
matrix element and due to its hermiticity we have $t_{i,j}^{ss^{\prime}}(\tau)=\left(t_{i,j}^{ss^{\prime}}(-\tau)\right)^{\star}$.
If each orbital $i$ is allowed to have its own quantization axis,
rotated with respect to the ``laboratory'' frame by the angle $\vartheta_{i}$
around the axis $\mathbf{n_{\mathrm{\mathit{i}}}}$ and there is a
pitch vector $\mathbf{Q}$, then $t_{i,j}^{ss^{\prime}}(\tau)$ becomes:
\[
\tilde{t}_{i,j}^{ss^{\prime}}(\tau)=\Omega_{su}^{\dagger}(i)t_{i,j}^{uu^{\prime}}(\tau)\Omega_{u^{\prime}u^{\prime\prime}}^{\phantom{\dagger}}(\tau)\Omega_{u^{\prime\prime}s^{\prime}}^{\phantom{\dagger}}(j),
\]
where we assume summation over repeated indices.

The Heisenberg exchange term has the most general form as follows:
\[
\sum_{\substack{i,j,r\\
\tau,\alpha,\beta
}
}J^{\alpha\beta}(i,j,\tau)S_{ir}^{\alpha}S_{jr+\tau}^{\beta}.
\]
The transformed matrix element $J^{\alpha\beta}(i,j,\tau)$ becomes:
\[
\tilde{J}^{\alpha\beta}(i,j,\tau)=\left(T^{\dagger}\left[\Omega(i)\right]J(i,j,\tau)T^{\phantom{\dagger}}\left[\Omega(\tau)\right]T^{\phantom{\dagger}}\left[\Omega(j)\right]\right)^{\alpha\beta},
\]
where the notation \emph{e.g.} $T\left[\Omega\right]$ signifies the
$3\times3$ vector rotation matrix by the angle $\vartheta$ around
the axis $\mathbf{n}$:
\begin{alignat*}{1}
T^{\phantom{\dagger}}\left[\Omega\right] & =T[\vartheta,\mathbf{n}]=\cos\vartheta I+\sin\vartheta\left(\begin{array}{ccc}
0 & \phantom{-}n_{z} & -n_{y}\\
-n_{z} & 0 & \phantom{-}n_{x}\\
\phantom{-}n_{y} & -n_{x} & 0
\end{array}\right)\\
 & +\left(1-\cos\vartheta\right)\left(\begin{array}{ccc}
n_{x}^{2} & n_{x}n_{y} & n_{x}n_{z}\\
n_{x}n_{y} & n_{y}^{2} & n_{y}n_{z}\\
n_{x}n_{z} & n_{y}n_{z} & n_{z}^{2}
\end{array}\right).
\end{alignat*}
Here $I$ is the $3\times3$ unit matrix.

\section*{References}

\bibliographystyle{elsarticle-num}
\bibliography{biblio}

\end{document}